\documentclass[printer]{aa}

\bibpunct{(}{)}{;}{a}{}{,}

\usepackage{graphicx}
\usepackage{txfonts}
\usepackage{soul}
\usepackage{txfonts}
\usepackage{color}
\usepackage{ulem}
\usepackage{lscape}
\usepackage{longtable}
\usepackage{rotating}
\usepackage{pdflscape}
\usepackage{subcaption}
\usepackage{float}
\usepackage{multirow}
\usepackage{booktabs}
\usepackage{caption}
\usepackage[export]{adjustbox}
\usepackage{subcaption}
\usepackage[version=4]{mhchem}
\usepackage{upgreek}
\graphicspath{{figures/}}

\def\cm3{cm$^{-3}$}

\def\2{$^{12}$CO}
\def\3{$^{13}$CO}
\def\8{C$^{18}$O}

\def\cm2{cm$^{-2}$}

\begin{document}

\title{Simple cyanides and formylium ions isotopologues in early star-forming molecular cores}

\author{ R. D. Taboada\inst{1,2,3,4}
\and S. Paron \inst{3}
\and N. C. Martinez \inst{3,4}
\and M. E. Ortega \inst{3}
}
\institute{CONICET - Universidad Nacional de Salta. Instituto de Investigación en Energía no Convencional. Salta, Argentina\\
             \email{rocio.taboada@exa.unsa.edu.ar}
\and Universidad Nacional de Salta. Facultad de Ciencias Exactas. Departamento de Matemática
\and CONICET - Universidad de Buenos Aires. Instituto de Astronom\'{\i}a y F\'{\i}sica del Espacio. Buenos Aires, Argentina
\and Universidad de Buenos Aires. Facultad de Ciencias Exactas y Naturales. Departamento de Física. Buenos Aires, Argentina
}

\offprints{R. Taboada}

   \date{Received <date>; Accepted <date>}

\abstract
{}{Understanding the chemistry related to the early stages of star formation is of great importance, as it is linked to the beginnings of the most complex chemistry in the interstellar medium. In this context,
we investigate the chemical behaviour of simple cyano-bearing molecules and formylium ions isotopologues in a sample of massive infrared-quiet molecular cores.}
{Using archive ALMA Band 7 data of 37 early molecular cores embedded in ATLASGAL clumps, we obtain abundances of HC$_3$N, H$^{13}$CN, HN$^{13}$C, H$^{13}$CO$^+$, and HC$^{17}$O$^+$. 
We used various statistical methods, including hierarchical clustering, to analyse the correlations between molecular abundances, ratios and temperature. }
{We find that  HN$^{13}$C, H$^{13}$CO$^+$, and HC$^{17}$O$^+$ abundances correlate positively with kinetic temperature, suggesting temperature-driven chemical regulation in young massive cores. A similar trend is observed for H$^{13}$CN, although the limited number of detections prevents a definitive conclusion. HC$_3$N abundances show no dependence on temperature within the 40--100 K range, suggesting a chemical steady state between gas-phase production and grain-surface depletion. Similarly, the H$^{13}$CN/HN$^{13}$C ratio, measured in only six regions, suggests no correlation with temperature, differing from findings at lower temperatures. Using a hierarchical clustering method based on abundance ratios, novel in astrochemistry, we identified chemically distinct core groups that align with thermal conditions. Additionally, we provide HC$^{17}$O$^+$ detections for 28 cores—a significant expansion of existing literature—and find evidence that H$^{13}$CO$^+$ transitions may have higher optical depths than commonly assumed. These results are important because characterizing the chemical state of early star-forming stages is essential for understanding the onset of the most complex chemistry.}
{}

\titlerunning{Simple molecules in early star-forming cores}
\authorrunning{R.D. Taboada et al.}


\maketitle

\section{Introduction}

Nowadays, it is well known that molecular clumps (with spatial scales of about 1 pc) can be resolved into smaller structures (spatial scales of a few tenths of pc and even smaller), known as molecular cores. This fragmentation is what leads to star formation \citep{beu25}. In other words, molecular cores represent the final stage of fragmentation of interstellar matter from which stars eventually form. Given that the richest chemistry in the interstellar medium (ISM) occurs in such cores (e.g. \citealt{herbst09,jor20,coletta20}), they are the natural astrochemical laboratories to investigate how complex molecules are formed in space and their relation with the star-forming processes. Indeed, the molecular matter is present at all spatial and temporal scales related to star formation \citep{jor20}. 

The rise of the chemical richness in molecular cores should be mainly correlated with temperature, as it is primarily driven by sublimation from dust-grain mantles, triggered by radiation and shocks from nascent stars \citep{beltran18}. As a consequence, hot molecular cores are highly rich in complex organic molecules (COMs).

An important aspect in the evolution of complex interstellar chemistry is the study of its origins, which must happen at cold interstellar regions associated with star formation. 
To best probe the initial chemical conditions at the earliest stages of star formation, it is necessary to study cold and
dense gaseous interstellar structures. In general, they are called starless cores and prestellar cores (e.g. \citealt{bergin07}). In low-mass star-forming regions, which are typically nearby and relatively isolated, starless cores are quite easy to identify and study, as seen for instance in the Perseus molecular cloud \citep{scibelli24}. In contrast, high-mass star-forming regions are more distant and deeply embedded, requiring an analysis of the inner structures within infrared-quiet massive clumps. Such clumps are the precursors to high-mass stars and clusters in an early evolutionary stage.

Molecular cores in the early stages of star-forming evolution are of high interest for constraining the early chemistry in the ISM. In previous works \citep{sulfur24,paron25}, we focused on the analysis of sulphur chemistry of a sample of 37 cores embedded in some of the most massive infrared-quiet molecular clumps from the APEX Telescope Large Area Survey of the Galaxy (ATLASGAL). Among several interesting findings, we found that the abundances of the analysed sulphur-bearing species are positively correlated with the temperature increase.  Using the same sample of regions and data set, in this work, we focus on the study of HC$_{3}$N, H$^{13}$CN, HN$^{13}$C, H$^{13}$CO$^{+}$, and HC$^{17}$O$^{+}$. 

The simplest cyanopolyyne, HC$_{3}$N, is one among the first organic molecules to be observed in the ISM \citep{turner71} and it seems to be ubiquitous in hot molecular cores \citep{bergin96,taniguchi16,duronea19}.  In addition, the HC$_{3}$N can trace the chemistry generated in the shocked gas by protostellar outflows \citep{hervias19,hoque25,lopez-g25}, in bow-shocks \citep{mendoza18}, and in the post-shocked gas \citep{busch24}.

Other simple molecules bearing a cyano group are hydrogen cyanide (HCN) and isocyanide (HNC). Such molecules and isotopologues have a linked chemistry, as pointed out many years ago by \citet{shcilke92}. Differences in their spatial distributions within a molecular structure can reflect the chemical conditions of the gas and the evolution in star formation. In that sense, it was proposed that the abundance ratio  $\mathrm{HCN/HNC}$ 
appears directly associated with the evolutionary stages of star formation \citep{jin15}. Moreover, using the intensity ratios between both isomers, it is possible to estimate the gas temperature in a range of 15 to 40 K \citep{hacar20,nai23,nai24}. However, as pointed out by \citet{nai24}, the HCN-HNC ratio as a kinetic temperature estimator should be used only after careful and dedicated revision of the spectra. For example, checking that no line, neither the main nor the hyperfine ones, display absorption features. Thus, using the isotopologues H$^{13}$CN and HN$^{13}$C, as generally optically thinner lines \citep{hirota98,jin15}, probably would yield more reliable correlations (see for instance \citealt{pazu22}).

The formylium ion (HCO$^{+}$) is also an ubiquitous molecular species across several interstellar environments. 
In dense, cold, and well-shielded regions impenetrable to far-ultraviolet photons, analysing this ion is crucial for studying the role that cosmic rays play ionizing molecular gas \citep{luo24}. At more evolved regions, where jets and outflows begin to be chemically dominant, an increase of HCO$^{+}$ is expected \citep{rawlings04,ortega12}, and it is even possible to trace molecular outflows using its emission \citep{sanchez13}. Therefore, the analysis of optically thinner HCO$^{+}$ isotopologues is also useful to investigate the chemistry of star-forming regions. It is worth noting that in the literature, there are very few studies based on observations of HC$^{17}$O$^+$ (e.g., \citealt{caselli02,ferrer22,mendoza23}). 

Studying these simple cyanides and formylium ion isotopologues is of significant interest in astrochemistry and astrophysics (e.g., \citealt{rb21}). Their detection and analysis in early-molecular cores are useful
for establishing an initial baseline for molecular abundances and the degree of isotopic fractionation. This foundational knowledge is crucial for understanding how chemistry evolves during the early stages of star formation.

\begin{table}[h]
\centering
\tiny
\caption{Analysed molecular lines.}
\label{transitions}
\begin{tabular}{llcccc}
\hline
\hline
Molecule & Transition & $\nu_{\rm rest}$ & E$\rm_{u}$ & $g_{u}$ & log($A_{ul}$)   \\  
         &            &   (GHz)          &   (K)      &         &                 \\ 
\hline
 & &   \\[-1.8ex]
HC$_{3}$N v=0    & J=37--36  & 336.5200  &  306.9  & 75  & -2.51   \\
H$^{13}$CN v=0   & J=4--3    & 345.3397  &  41.4   & 9  &  -2.72   \\
HN$^{13}$C       & J=4--3    & 348.3402  &  41.8   & 9  &  -2.79   \\
H$^{13}$CO$^{+}$ & J=4--3    & 346.9983  &  41.6   & 9  &  -2.48   \\
HC$^{17}$O$^{+}$ & J=4--3    & 348.2110  &  41.7   & 9  &  -2.47   \\
\hline
 & &   \\[-1.8ex]
\multicolumn{6}{l}{\tiny Note: Parameters extracted from the Splatalogue Database for Astronomical}\\ 
\multicolumn{6}{l}{\tiny Spectroscopy using the Cologne Database for Molecular Spectroscopy} \\
\multicolumn{6}{l}{\tiny (CDMS; \citealt{muller05}). }\\ 
\end{tabular}
\end{table}

\section{Data and source sample}
\label{sectdata}

This work was carried out with the same data from the Atacama Large Millimeter Array (ALMA) as used in \citet{sulfur24}. Such a data set (project 2017.1.00914.S; PI: Csengeri) was retrieved from the ALMA Science Archive\footnote{http://almascience.eso.org/aq/}. This observing project consisted of observations in Band 7 toward massive infrared-quiet clumps in the inner Galaxy. They are ATLASGAL sources containing dense fragments, such as molecular cores in the early stages of star formation. Having analysed sulfur-bearing molecules in our former work, in this 
case we selected the molecular lines listed in Table\,\ref{transitions} for the characterisation of the chemistry in the cores. Although these data were described in our previous paper, we briefly mention some of their main characteristics for completeness. The observed
frequency range and the spectral resolution are 333.3–349.1 GHz and 1.1 MHz. The angular resolution is about 3$\farcs$5. The continuum and line (each 10 km\,s$^{-1}$) sensitivity are about 1.2 and 30~mJy beam$^{-1}$. It is important to note that the data passed the Quality Assurance Level 2, ensuring reliable calibration for science-ready data. 

The 37 ATLASGAL sources and the coordinates, LSR velocity, and distance of the main cores embedded within them were presented in Table\,A.1 of Appendix\,A in \citet{sulfur24}. It is important to mention that kinetic temperatures (T$_{\rm k}$) of each core were estimated from the rotational diagram method (RD) performed from the CH$_{3}$OH 7(1,7)-6(1, 6)++ (335.582 GHz, E$_{\rm u}$ = 78.97 K) and 12(1,11)-12(0,12)-+ (336.865 GHz, E$_{\rm u}$ = 197.07 K) lines. Appendix\,C in \citet{sulfur24} explains the procedure, and the obtained rotational temperatures (T$_{\rm CH_{3}OH}$) are presented in Table\,D.2 of the mentioned work. It is worth noting that since the mentioned CH$_3$OH lines are the only ones shared by all cores, they were used uniformly throughout the previous work to ensure consistency across the entire sample of cores. In that work and here, it was assumed that T$_{\rm k}$ is equal to T$_{\rm CH_{3}OH}$, which is utilized as the thermal characteristic of each core. The validity of this equality is further discussed in the next Section.

\begin{figure*}[h!]
\centering
\includegraphics[width=6cm]{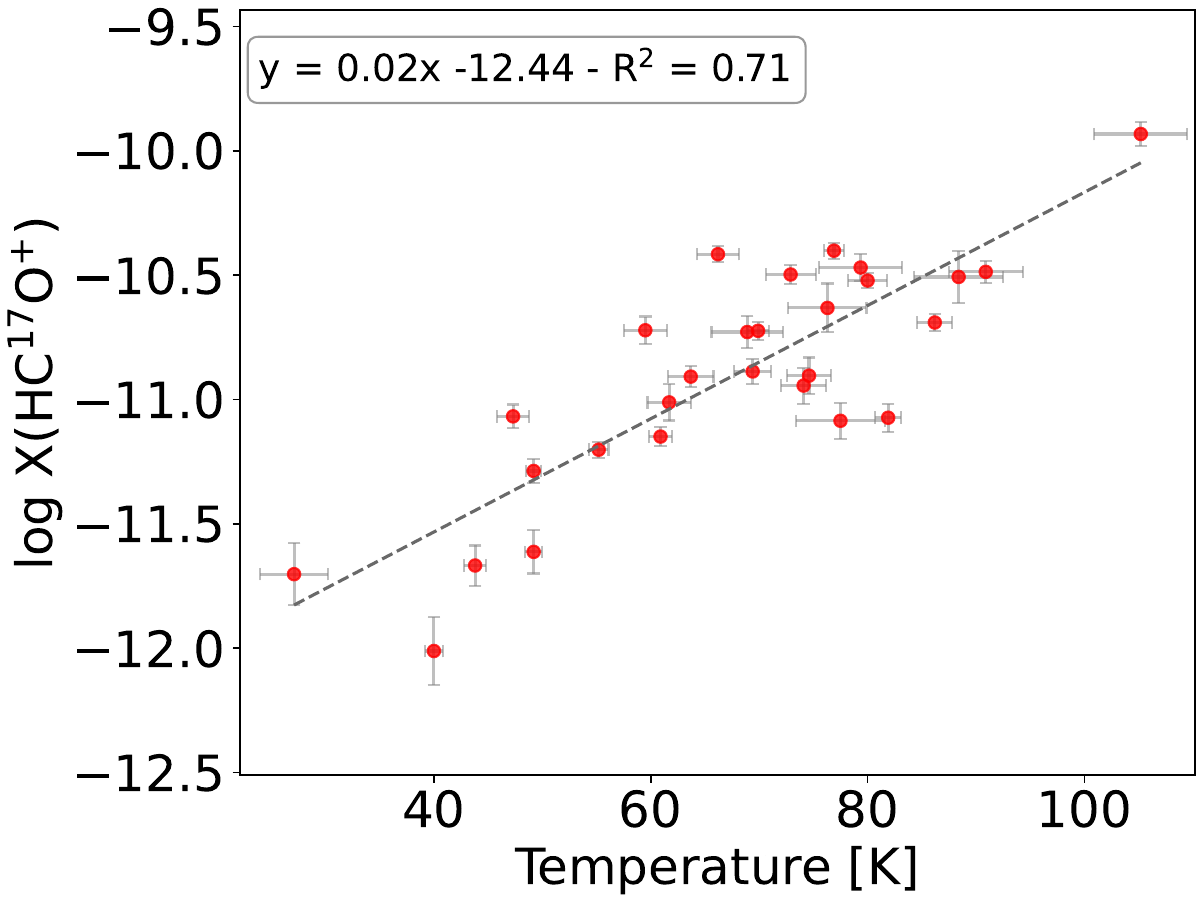}
\includegraphics[width=6cm]{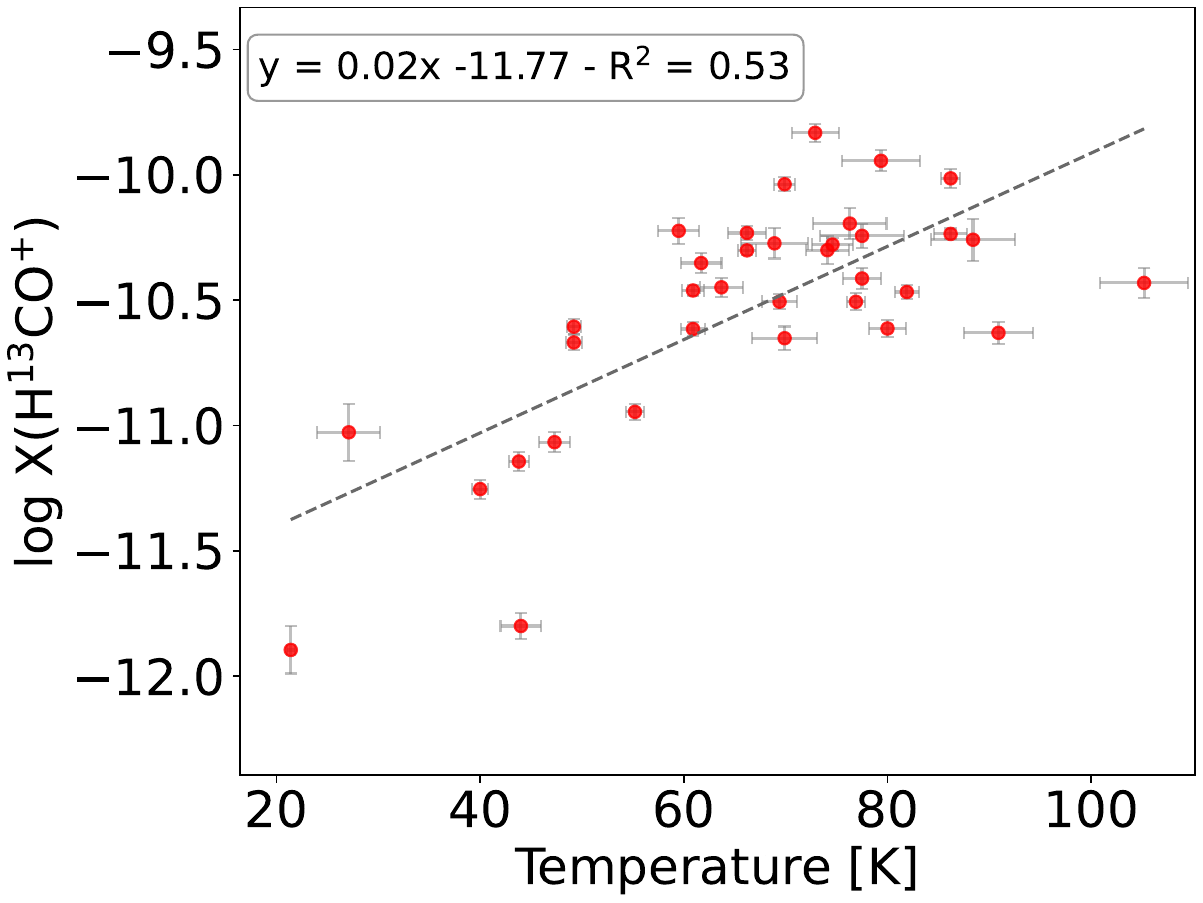}
\includegraphics[width=6cm]{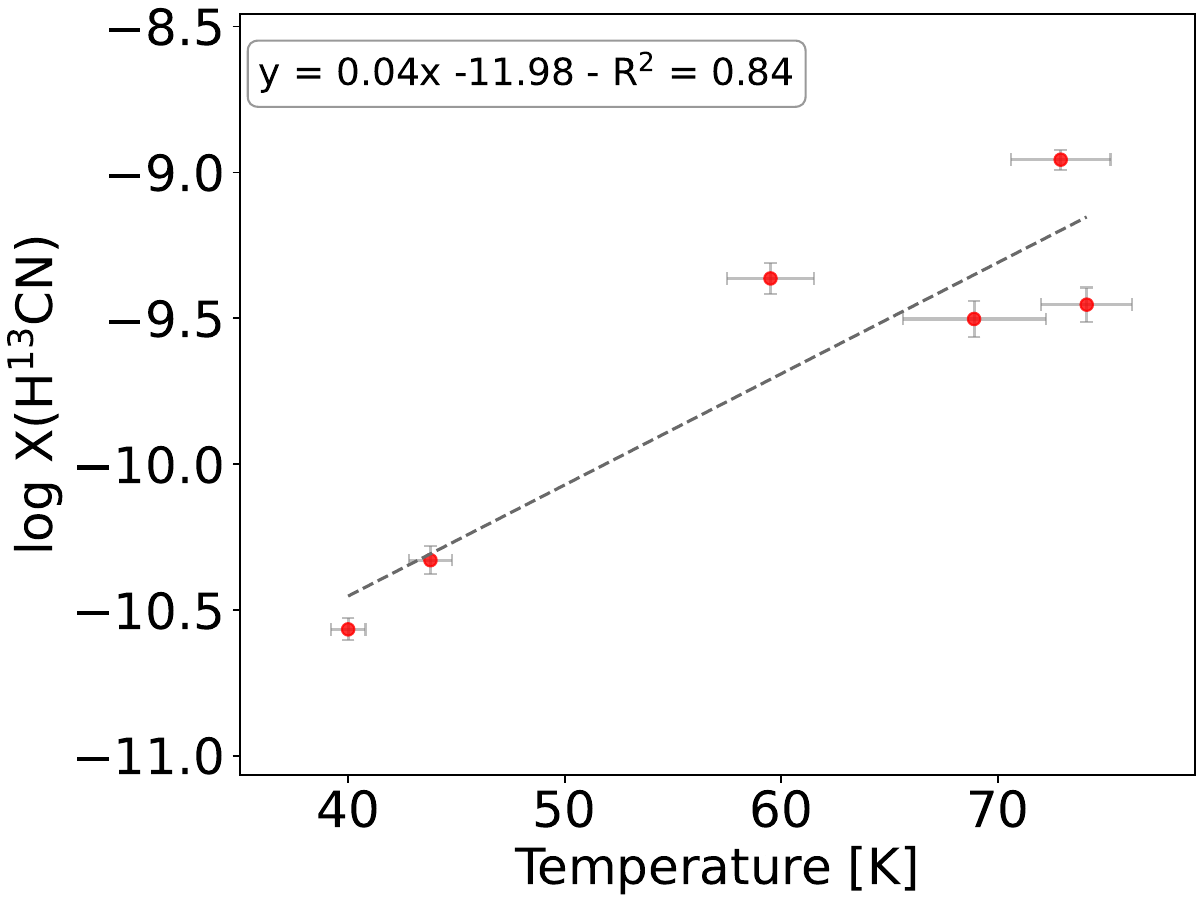}
\includegraphics[width=6cm]{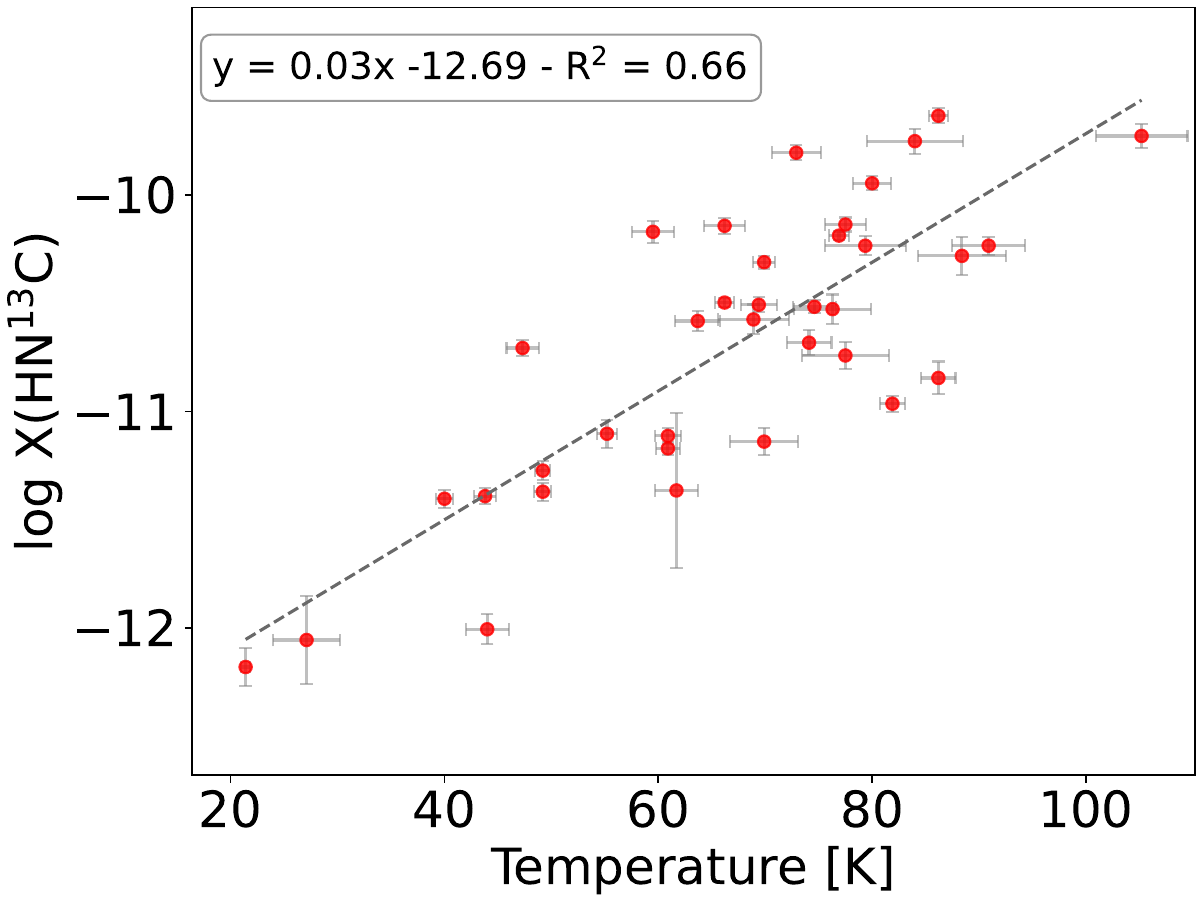}
\includegraphics[width=6cm]{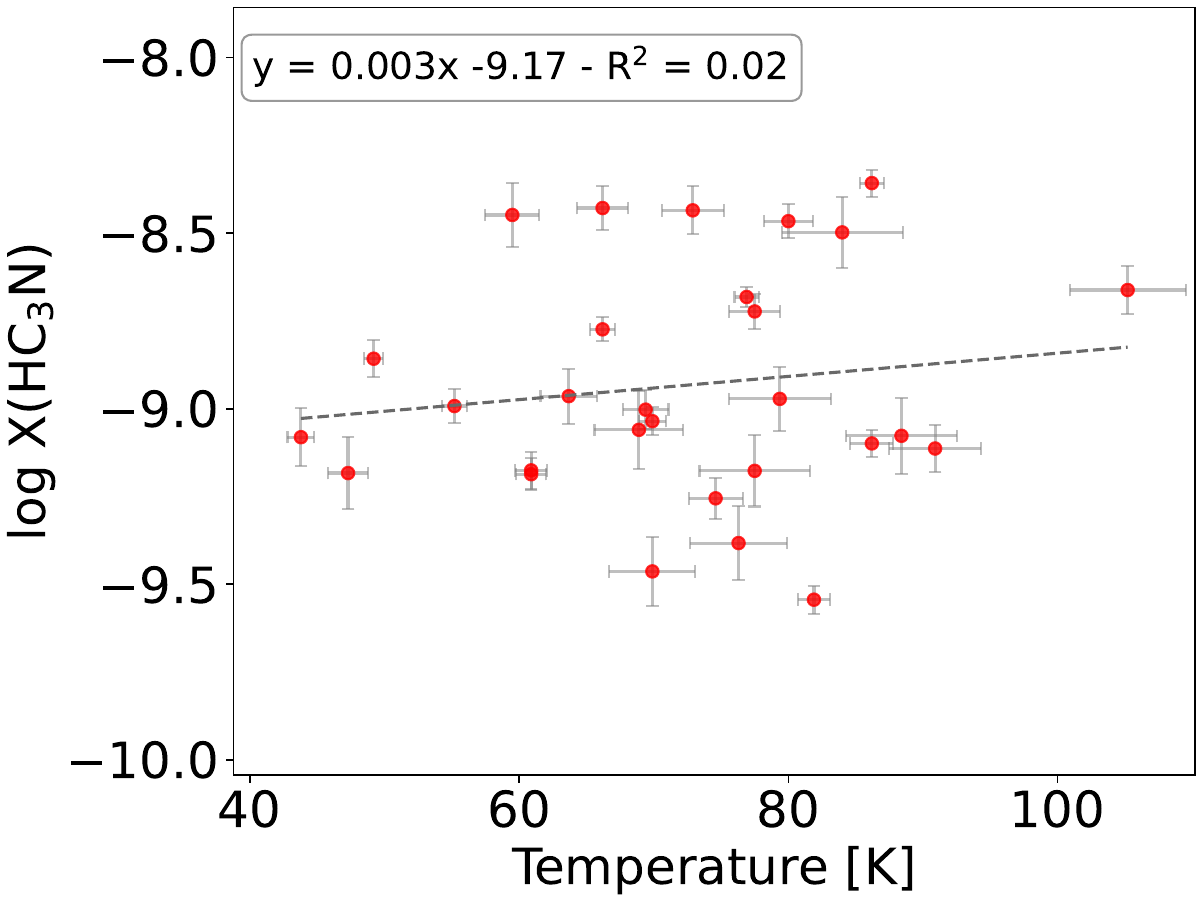}
\caption{Logarithmic molecular abundances
as a function of kinetic temperature. Linear fits are indicated by dashed grey lines; the best-fit parameters are listed in the top-left corner of each panel. Kinetic temperatures were extracted from \citet{sulfur24} (see text). Errors in the slopes obtained from the linear fitting process are between 10 and 20\%.}
\label{XvsT}
\end{figure*}

\section{Results}
\label{results}

Spectra from each core embedded in each ATLASGAL source were extracted at their continuum peak position using a beam-size region.  In all cases, the peaks of the molecular emission of the analysed species match those of the continuum emission. Gaussian fittings to each line were applied to obtain the intensity peaks, the full width half maximum (FWHM) $\Delta$v, and the integrated line intensity (i.e., the line flux, W). As an illustrative example, in Appendix\,\ref{spectraexample} we present the spectra obtained toward region  G310.1440+0.7592 (core\,22) (see Fig.\,\ref{fig:spectra_G310.1440+0.7592}). All the parameters obtained from the Gaussian fittings are presented in Tables\,\ref{tab:hc3n}, \ref{tab:h13cN-hc13N}, and \ref{tab:h13co-hc17o} in Appendix\,\ref{append0}.   
It is important to note that none of the analysed lines show signs of absorption, flattening, or signatures of spectral wings. As already noted in our previous work \citep{sulfur24}, it is important to remark that the spectra do not exhibit a large population of molecular lines. This is consistent with the nature of the sources, cores with temperatures below 100 K, and it facilitated line identification. 

In the case of H$^{13}$CN J=4--3, unfortunately, we could measure only a few lines. This is because many spectra covering this line are truncated near 345.3 GHz, leaving out the H$^{13}$CN emission. Regardless, those points  can be useful for exploring some kind of trend. It is important to note that the line consists of hyperfine components, not resolved within the spectral resolution of the observations. Hence, the H$^{13}$CN J=4--3 emission was fitted with a single Gaussian function.

To estimate the column densities of each molecular species, assuming local thermodynamic equilibrium (LTE) and following \citet{artur23}, we first derived the column densities of the
upper levels, N$_{\rm u}$ (in cm$^{-2}$), of each transition from

\begin{equation}
{\rm N_u = 2375 \times 10^6~\left(\frac{W}{Jy~beam^{-1}~km~s^{-1}}\right)~\left(\frac{s^{-1}}{A_{ul}}\right)~\left(\frac{arcsec^2}{\uptheta_{maj}\times\uptheta_{min}}\right)}
\label{coldensEq}
\end{equation}

\noindent where W is the measured line flux in Jy beam$^{-1}$ km s$^{-1}$, A$_{\rm ul}$ is the line Einstein coefficient (in s$^{-1}$) (see Table \ref{transitions}), and  $\rm \uptheta_{maj}$ and $\rm \uptheta_{min}$ are the half-power beam widths along the major and minor axes, respectively.

Then, the total column density of each molecule was obtained from

\begin{equation}
{\rm N_{tot} = \frac{N_u~Q(T_{ex})~exp(E_{u}/T_{ex})}{g_u}}
\label{coldensEq}
,\end{equation}

\noindent where ${\rm Q(T_{ex})}$ is the rotational partition function, ${\rm T_{ex}}$ is the excitation temperature (in K), $g_{\rm u}$ the upper-level degeneracy, and E$_{\rm u}$ (in K) the upper energy of the line transition.  All these parameters, except Q, are shown in Table\,\ref{transitions}. 
Following LTE considerations, ${\rm T_{ex}}$ was assumed to be equal to the kinetic temperature ($\rm T_k$) of the gas in each core. The values of $\rm T_{k}$ were obtained in \citet{sulfur24} from the rotational diagram method applied to CH$_3$OH transitions for each core. The temperature-dependent partition function was approximated from the CDMS catalog via the Splatalogue database, adopting the tabulated value for the temperature closest to the measured $\rm T_k$ of each core.
Obtained column densities and their respective errors are presented in Table\,\ref{coldens}.

As mentioned in Sect.\,\ref{sectdata}, in \citet{sulfur24} two methanol lines were used uniformly across the entire core sample to obtain temperatures from RDs. Although a two-point RD prevents a formal statistical linear fit, the reliability of the derived T$_{\text{rot}}$ is physically well-constrained. First, the chosen transitions cover a suitable energy range ($\Delta$E$_{\rm up} \approx$ 120 K; from E$_{\rm u}$ about 80 to 197 K) that is sufficiently wide to minimize sensitivity to calibration errors, yet narrow enough to probe the same gas component under LTE conditions. Second, as stated in \citet{sulfur24} it was verified that the slopes in the RDs are unaffected by opacity effects, rendering these two lines a reliable proxy for the rotational temperature. 
Regarding the assumed equality T$_{\rm rot}$=T$_{\rm k}$=T$_{\rm ex}$, it is important to mention that under LTE, the rotational temperature usually characterizes the kinetic temperature of the region. This equality holds valid in high-density environments, such as the analysed molecular cores, where the local density of gas exceeds the critical density of the observed transitions. In this regime, and moreover taking into account that the regions are catalogued as infrared-quiets, collisional processes dominate over radiative decay, thermalizing the molecular energy levels to the kinetic energy of the surrounding gas \citep[see][]{goldsmith99,mangum15}.

\begin{figure*}[h!]
    \includegraphics[width=6.2cm]{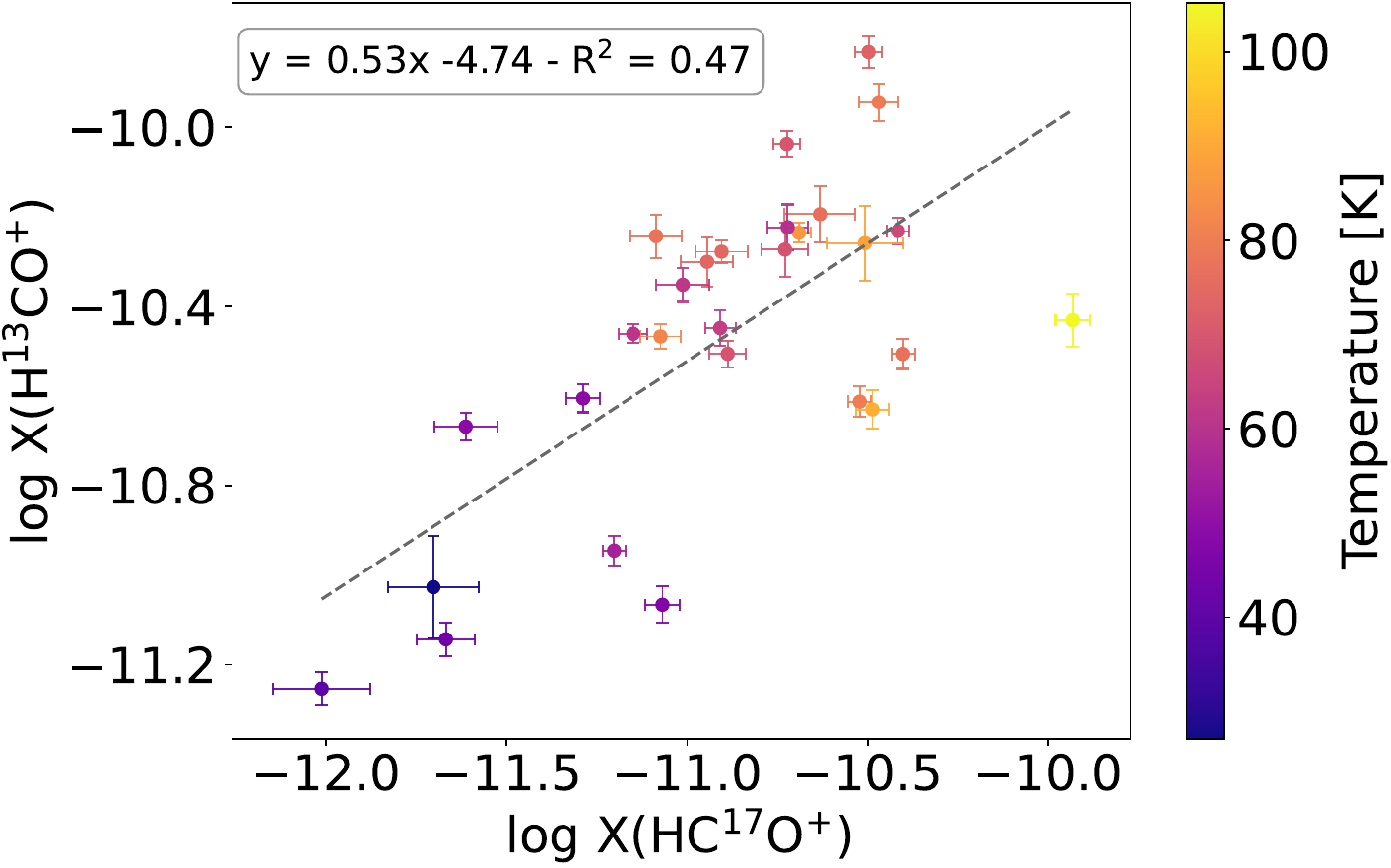}
    \includegraphics[width=6.2cm]{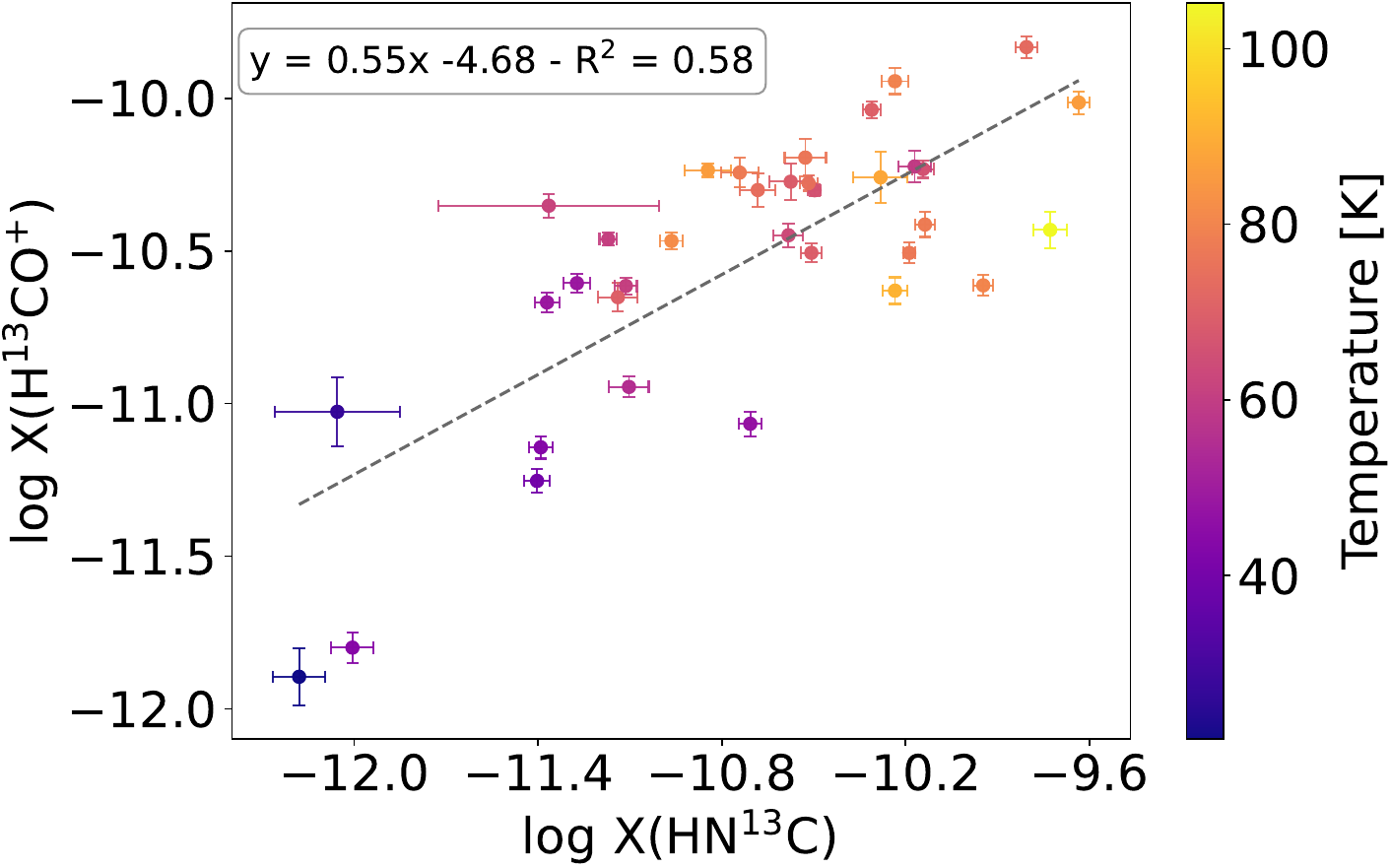}
    \includegraphics[width=6.2cm]{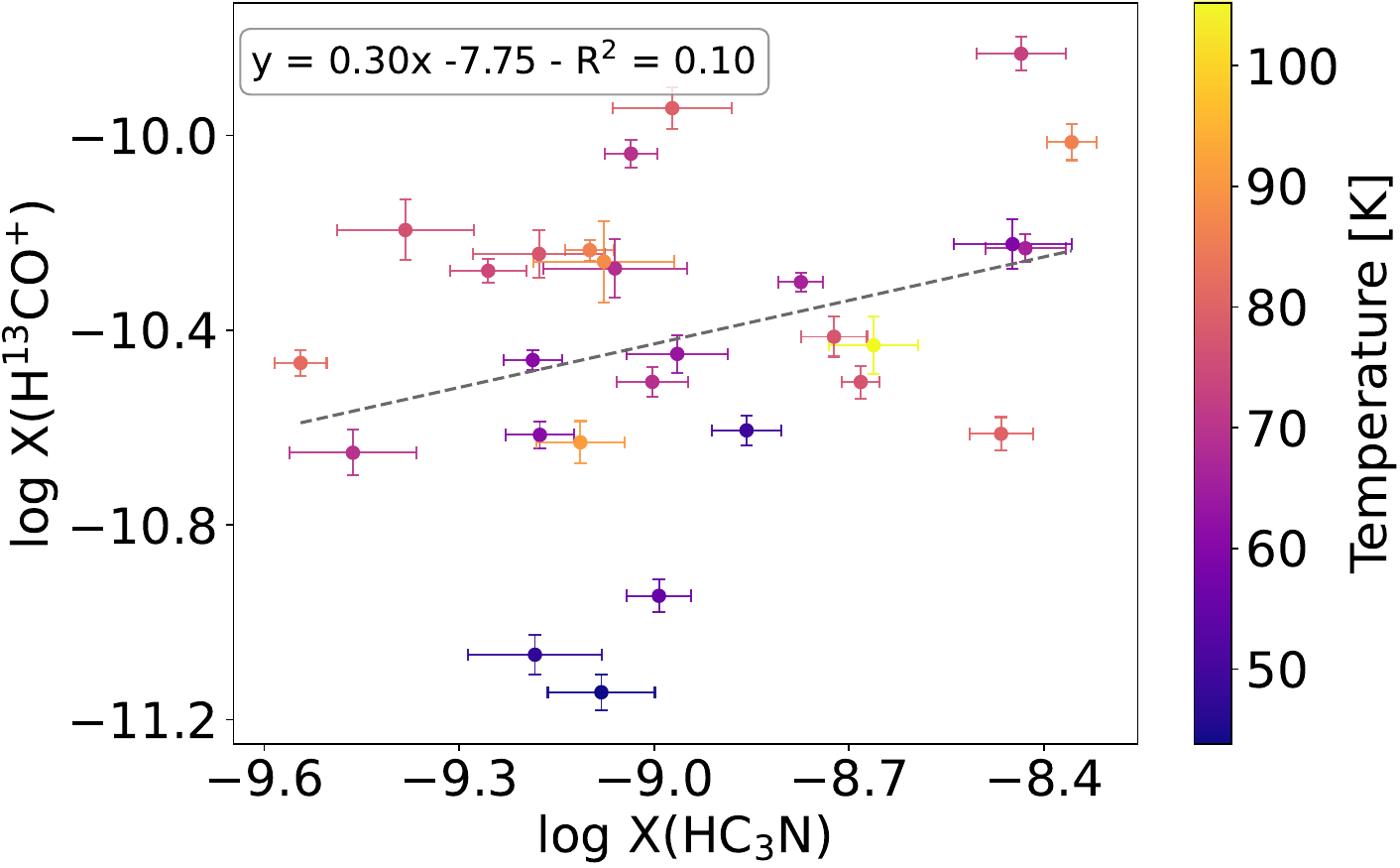}
    \vspace{0.2cm}
    \includegraphics[width=6.2cm]{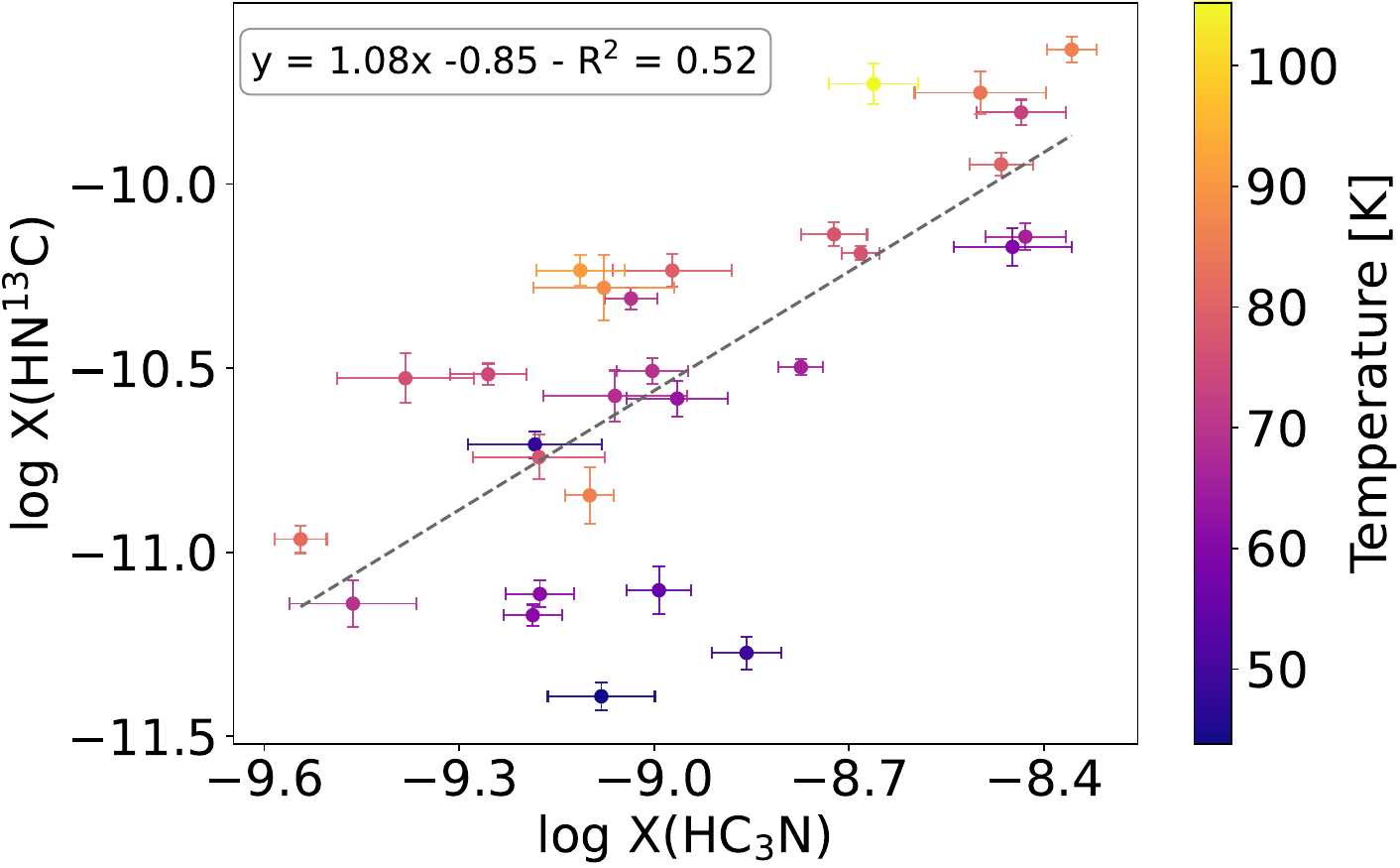}
    \includegraphics[width=6.2cm]{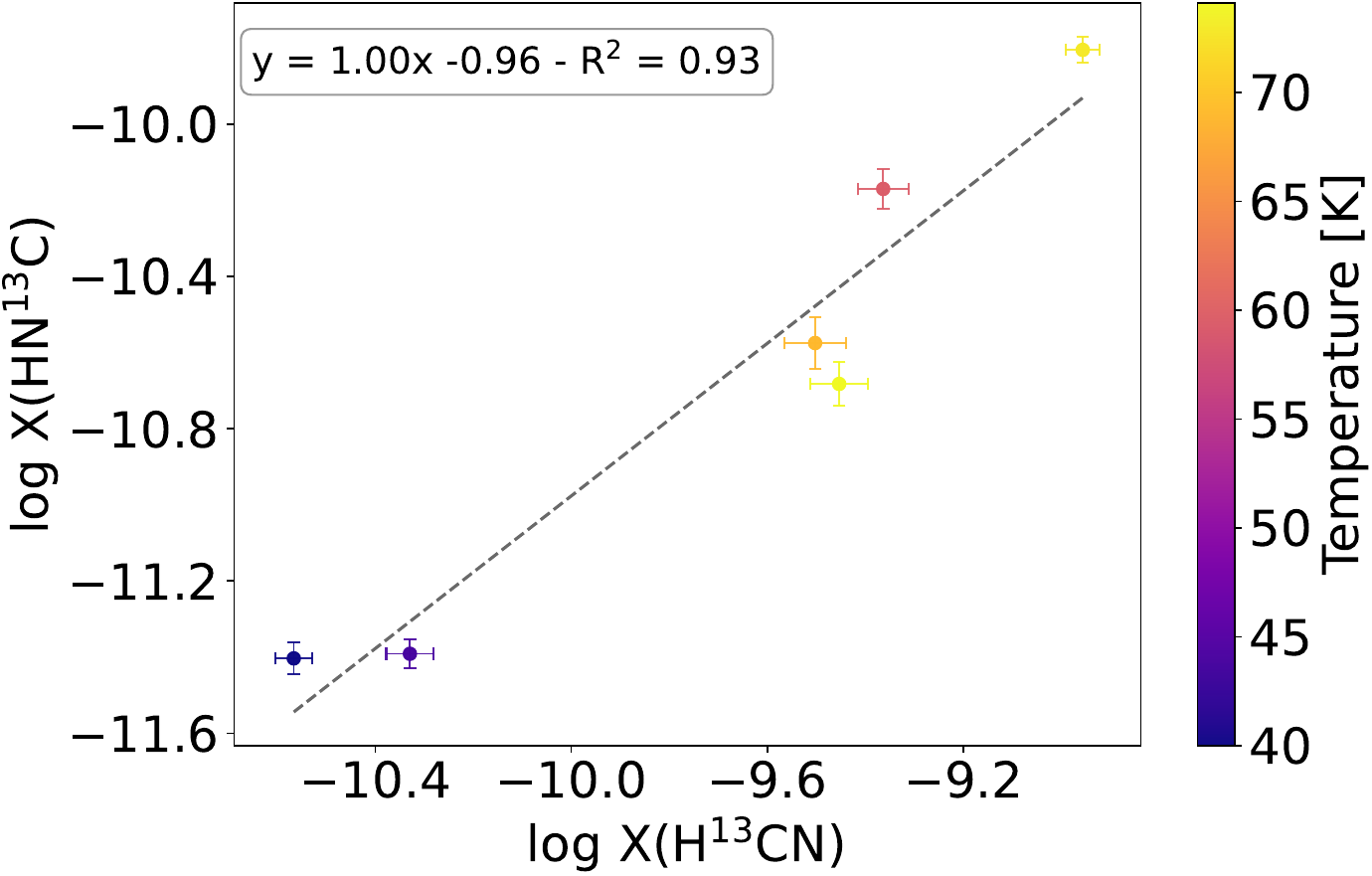}
    \includegraphics[width=6.2cm]{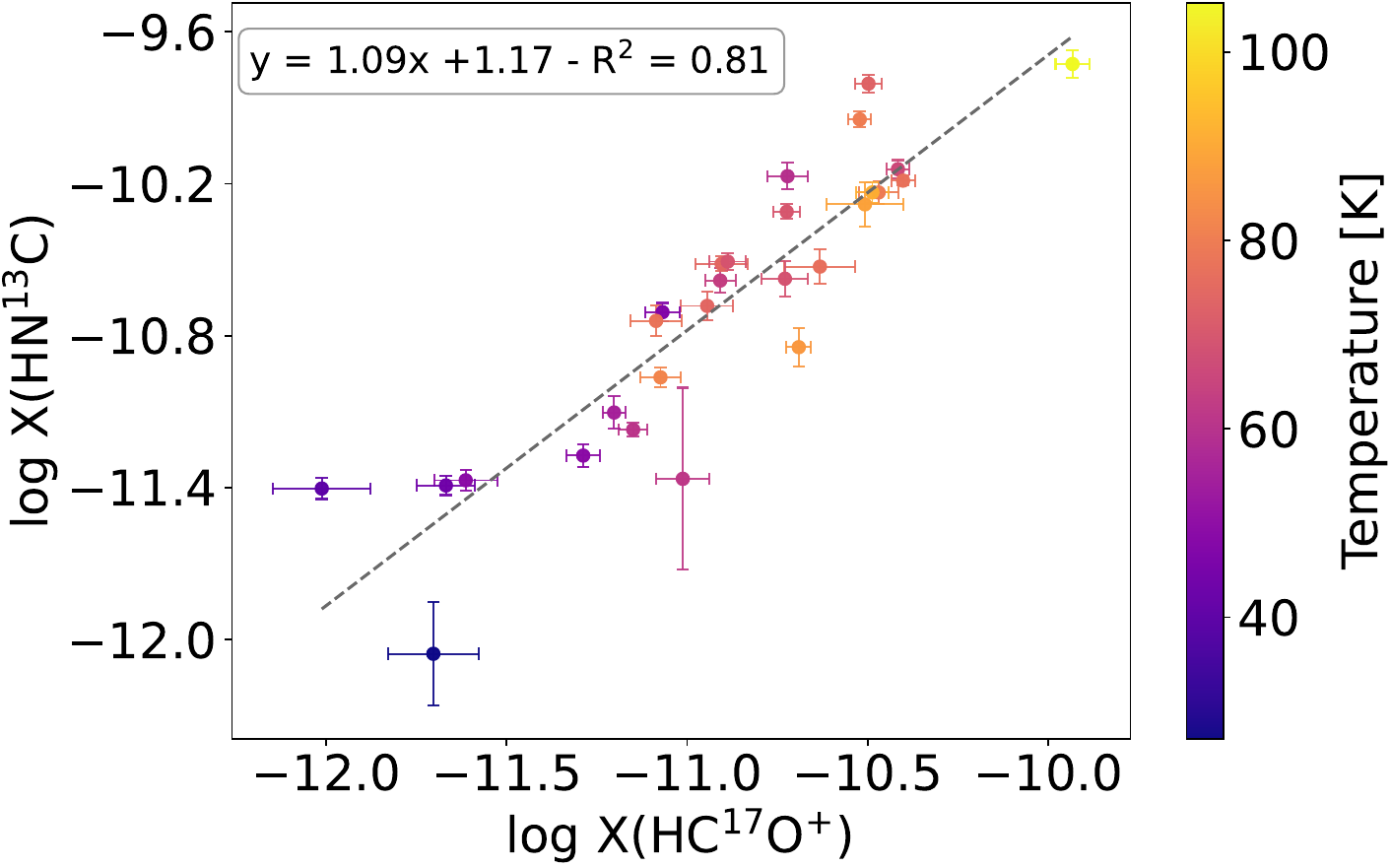}
    \vspace{0.2cm}
    \includegraphics[width=6.2cm]{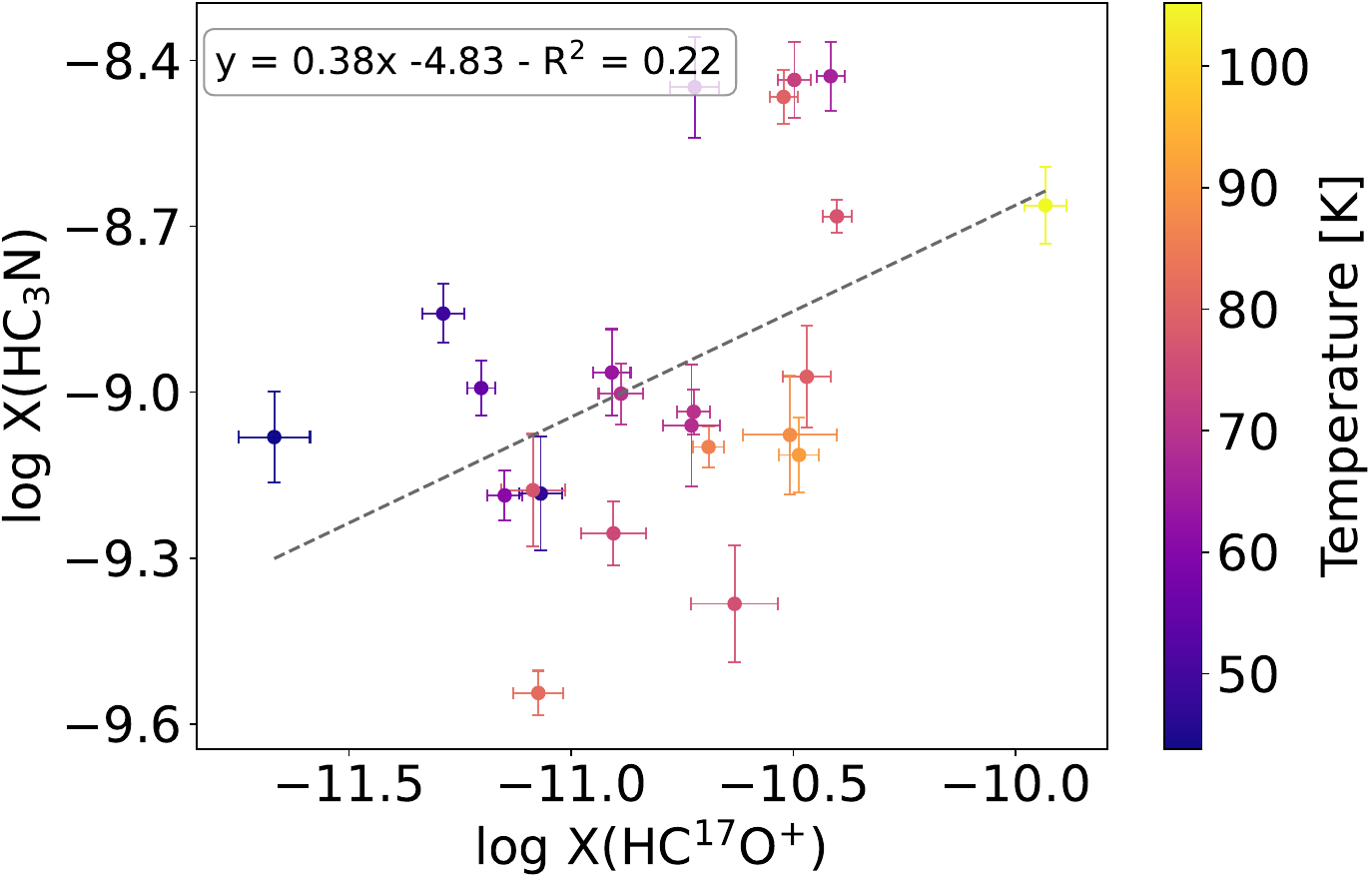}
    \includegraphics[width=6.2cm]{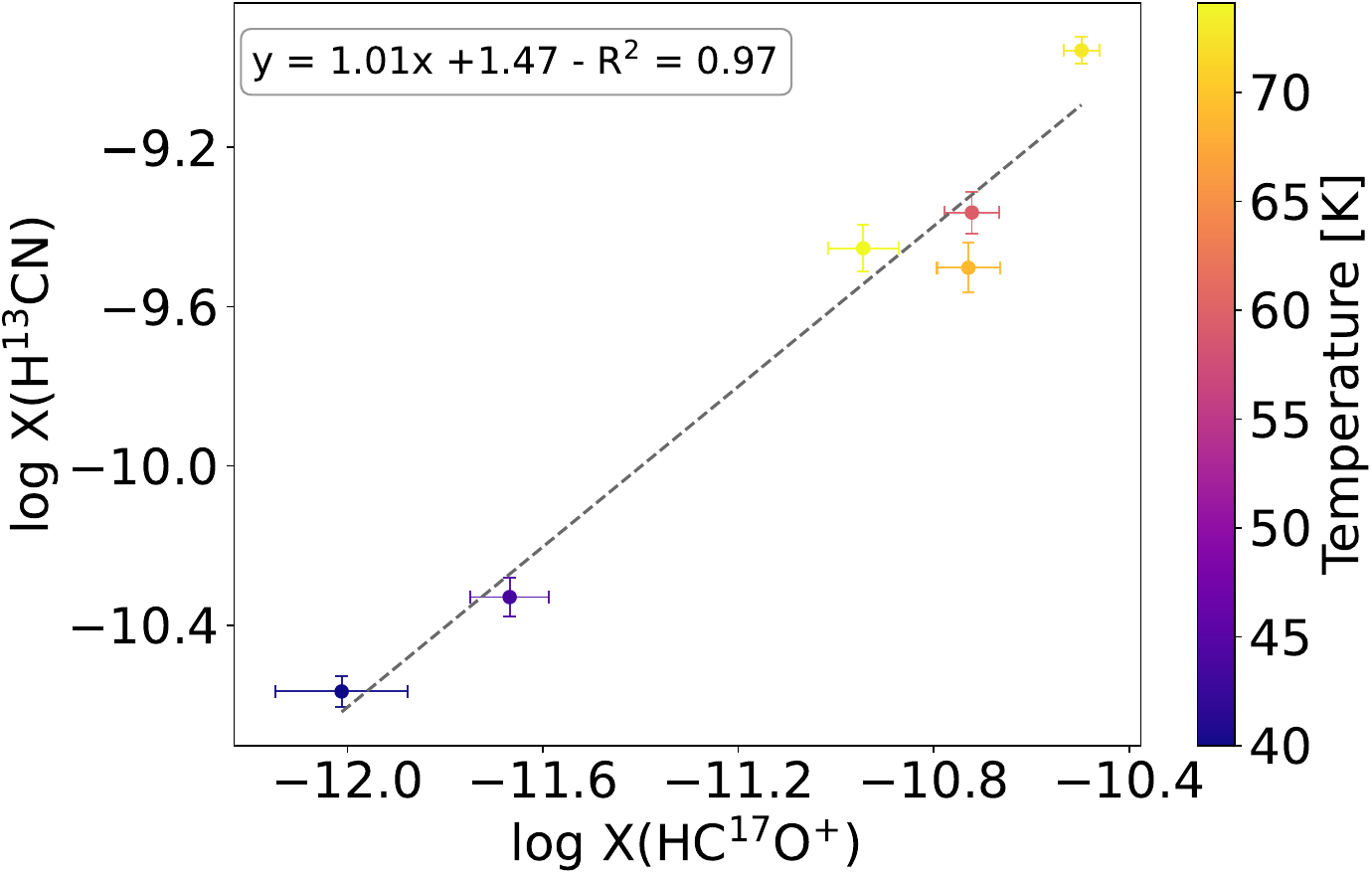}
    \includegraphics[width=6.2cm]{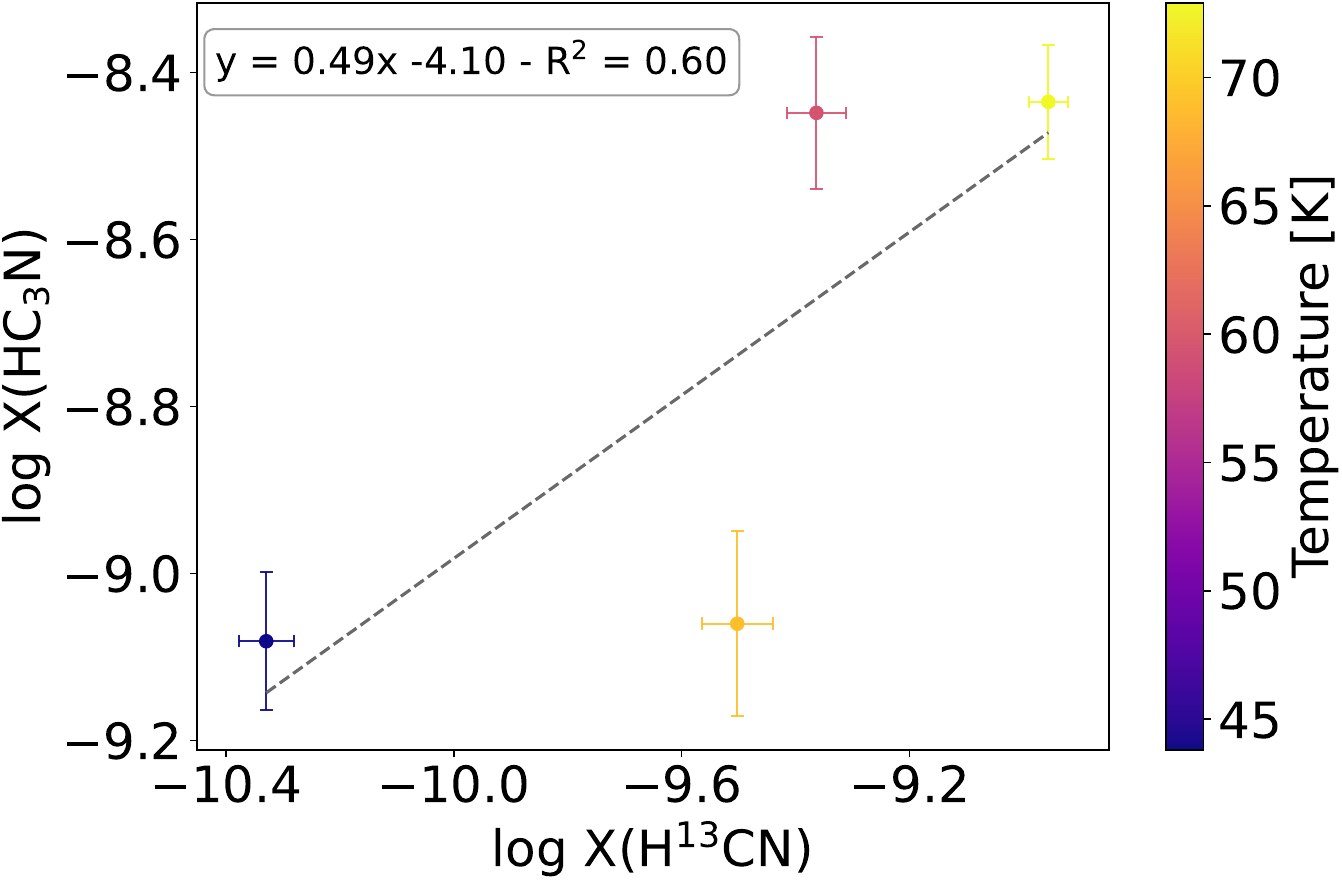}
    \vspace{0.2cm}
  \begin{minipage}[c]{6cm}
       \includegraphics[width=6.1cm]{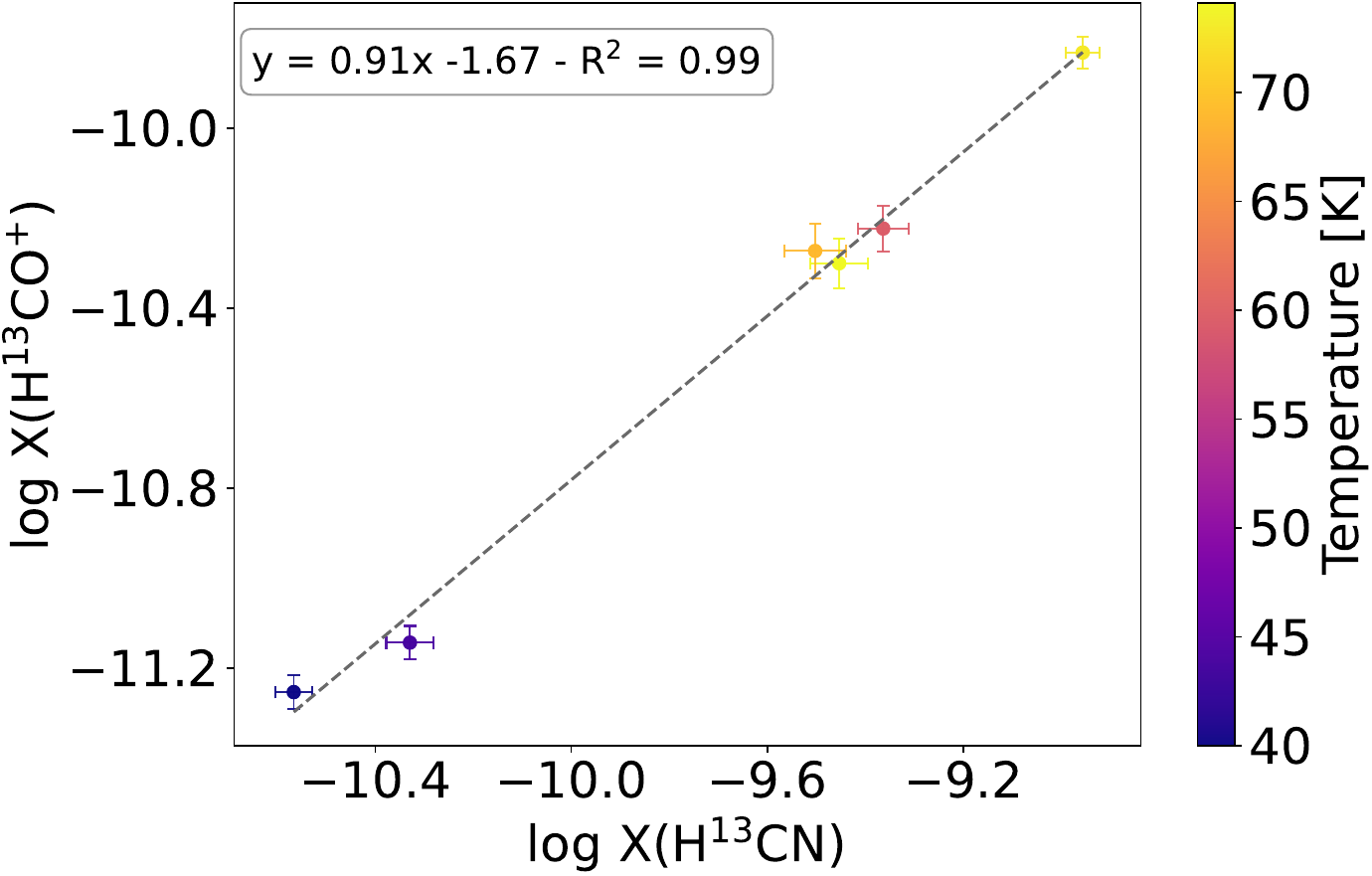} 
    \end{minipage}
    \hspace{0.85cm}
    \begin{minipage}[c]{10cm}
        \vspace{-0.6cm}
        \caption{Comparison between molecular abundances of all analysed species presented in logarithmic scale. Dashed lines correspond to linear fits, and the fit parameters are indicated in each panel. Colours indicate the kinetic temperature of the cores.  Errors in the slopes obtained from the linear fitting process are between 10 and 20\%.}
        \label{fig:abundancias}      
    \end{minipage}
\end{figure*}

\begin{figure}[h]
\centering
\includegraphics[width=8cm]{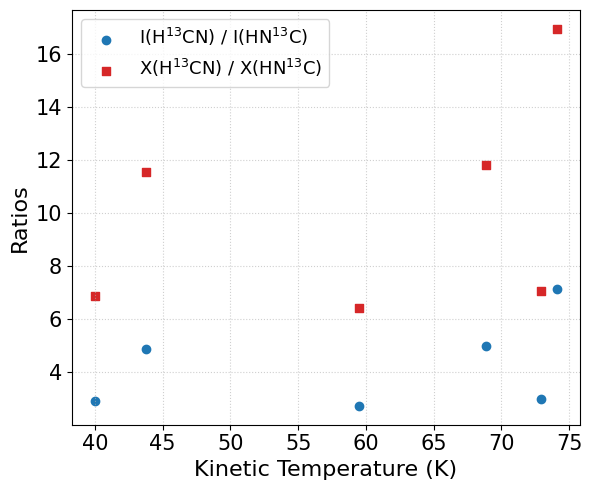}
\caption{ H$^{13}$CN/HN$^{13}$C integrated line ratios (blue circles) and abundance ratios (red squares) versus kinetic temperature from the few cores having observations with the H$^{13}$CN J=4--3 line.}
\label{isomers}
\end{figure}

By defining the abundance of a molecular species as X(molecule) = N(molecule)/N(H$_{2}$), this parameter was calculated for each species at each core, and they are presented in Table\,\ref{tab:abund}. The N(H$_{2}$) values were calculated in \citet{sulfur24} (see their Table\,D2).  

Comparisons between the molecular abundances derived for each species and the kinetic temperature are presented in Fig.\,\ref{XvsT}. Figure\,\ref{fig:abundancias} presents correlations between different abundances colour-code by temperature. In both figures, linear fits are provided in each panel to investigate potential underlying trends.

For the six sources with H$^{13}$CN measurements, the relationship of the $\rm H^{13}CN/HN^{13}C$ integrated emission and abundance ratios with respect to temperature is presented in Fig.\,\ref{isomers} (upper and lower panels, respectively). It is important to note that the values of both quantities are proportional and exhibit the same behaviour with temperature. This suggests that the lines are indeed optically thin and their emission is co-spatial.

\subsection{Molecular statistical analysis}
\label{stats}

Motivated by the results obtained above, which suggested correlations both among the molecular abundances and with the temperature (examined via simple linear fits), this section expands our statistical analysis to identify alternative correlation patterns.

\begin{figure}[h]
\centering
\includegraphics[width=9cm]{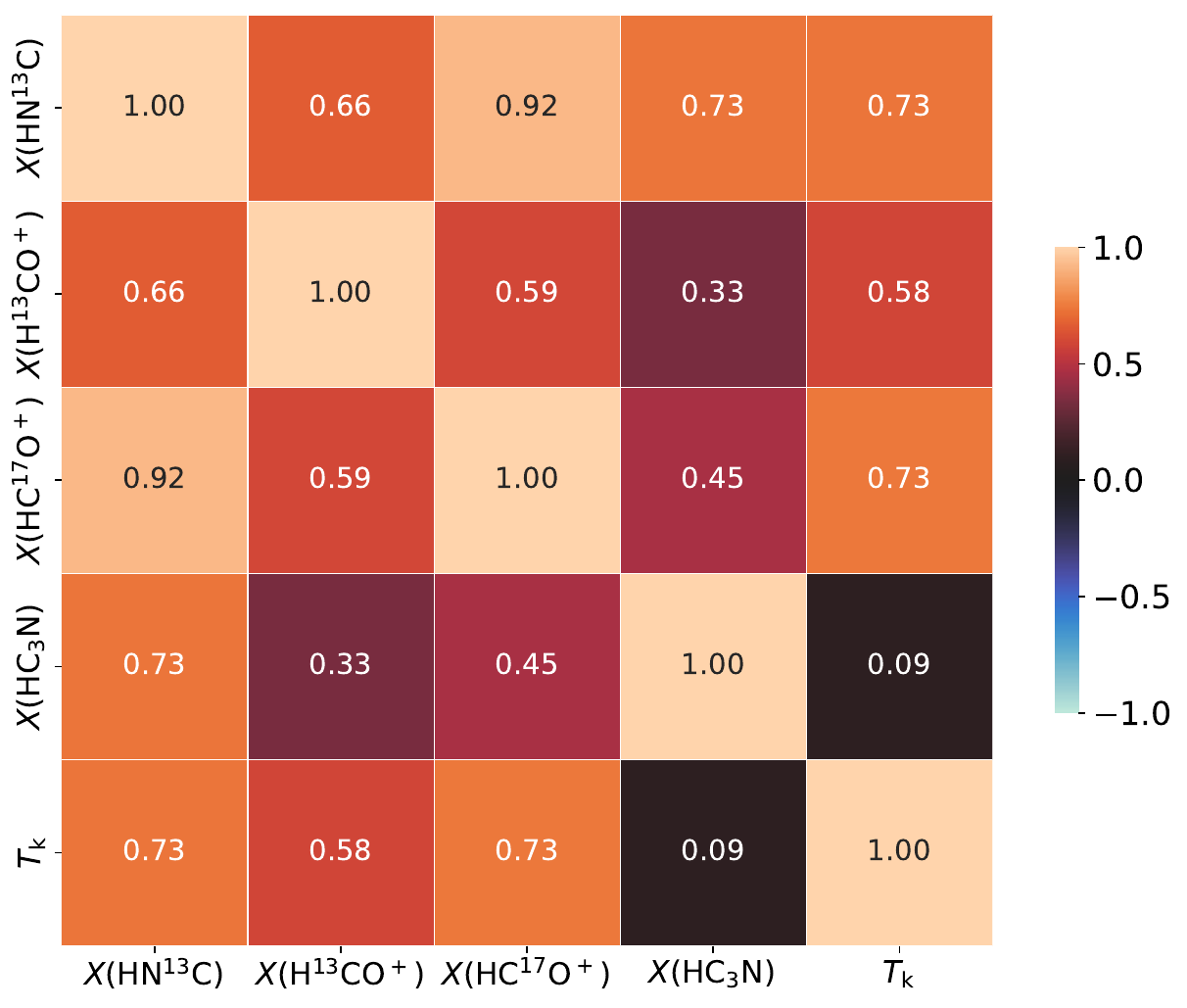}
\caption{Spearman correlation matrix between the logarithmic molecular abundances and kinetic temperature.}
\label{fig:corr_matrix}
\end{figure}

We evaluated the monotonic relationships among the molecular abundances and the kinetic temperatures by computing a Spearman rank correlation matrix \citep{spearman1904} as it was done for instance in \citet{rb21} with a larger set of molecular abundances. The non-parametric coefficient in the Spearman matrix ($\rho$) allows us to assess possible monotonic trends without assuming linearity. In addition, we evaluated the statistical significance of the correlations, testing the null hypothesis, which posits that there is no monotonic association between the variables in the population ($\rho = 0$); that is, determining the p-value. Low p-values indicate that the observed correlations are unlikely to arise by chance under the null hypothesis.
The abundance of H$^{13}$CN is excluded from this analysis due to the limited number of data points available. The obtained Spearman correlation matrix
is presented in Fig.\,\ref{fig:corr_matrix}.

This analysis shows that the abundance of HC$_3$N presents a negligible correlation with the temperature ($\rho = 0.09$, with a p-value $=$ 0.65), indicating the absence of a statistically meaningful monotonic dependence. In contrast, the other analysed species exhibit positive correlations with T$_{\rm k}$. The relations among the abundances also show positive correlations, and in general they have Spearman coefficients above 0.5. We verified that these correlations are statistically significant (p $<$ 0.01), meaning that they are unlikely to arise from randomness.  

We employed a hierarchical clustering analysis as a visual and statistical tool to investigate chemical similarities among the molecular cores.
Hierarchical clustering has been used in astronomy to identify structures in multivariate datasets and to explore similarities between sources \citep{boesso18,yu2022hierarchical}. 
This procedure quantifies similarities between data according to a predefined metric and represents them through a dendrogram, which can reveal hierarchical relationships. In this work, the similarity between sources is quantified using the Euclidean distance computed on the standardised abundance ratios.

To search for chemical similarities among the molecular cores based on the analysed molecular species, the clustering analysis was performed using the abundance ratios  HN$^{13}$C/HC$_{3}$N, H$^{13}$CO$^{+}$/HC$_{3}$N, and HC$^{17}$O$^{+}$/HC$_{3}$N. We decided to analyse the abundances relative to HC$_{3}$N following previous studies that identify this molecule as an early-type species. For instance, \citet{tani19} studied the N$_{2}$H$^{+}$/HC$_{3}$N abundance ratio as a chemical evolutionary indicator of massive cores. We follow a similar approach, also using ions and incorporating HN$^{13}$C.The use of abundance ratios relative to HC$_3$N reduces the influence of overall abundance differences between sources, preventing the clustering analysis from being dominated by absolute abundance levels. This approach is also motivated by the fact that HC$_3$N shows at most a very weak dependence on temperature.

\begin{figure}[t]
\centering
\includegraphics[width=\linewidth]{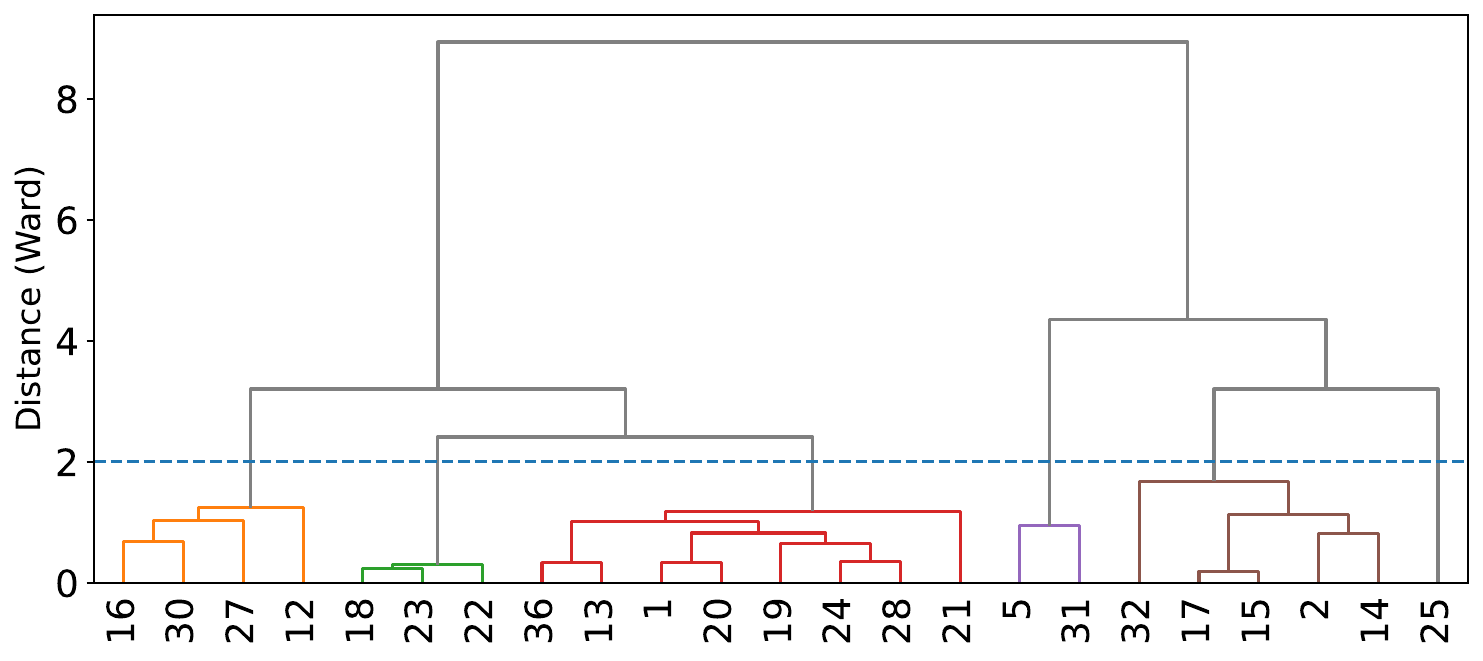}
\vspace{0.2cm}
\includegraphics[width=\linewidth]{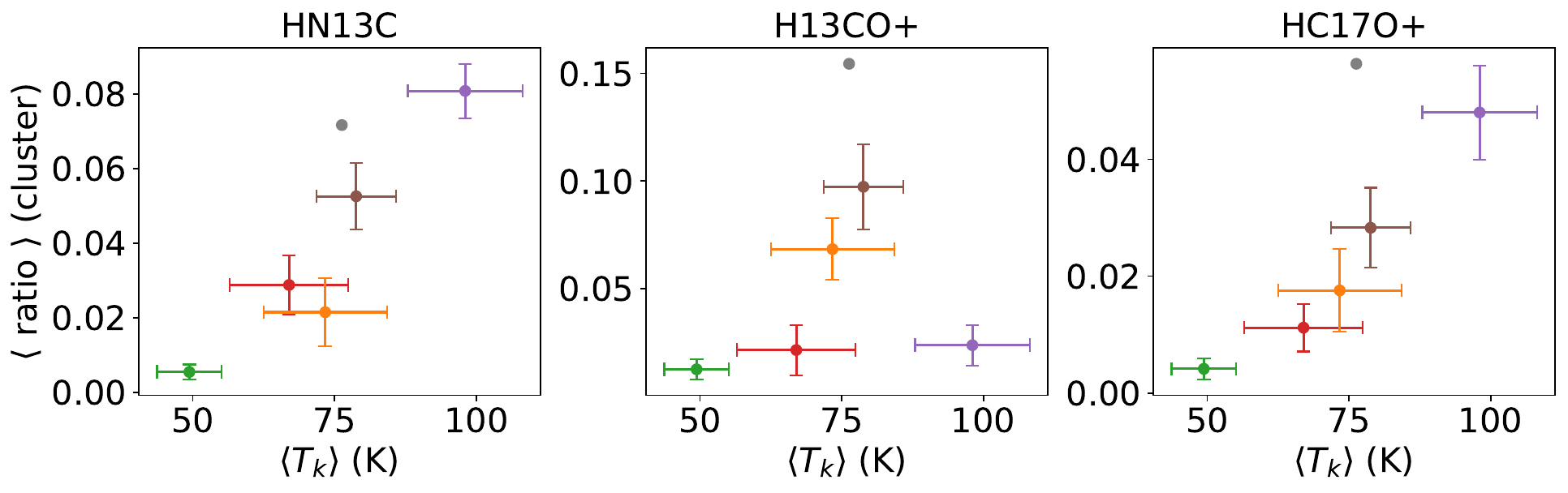}
\caption{
Hierarchical clustering of the molecular cores based on the abundance ratios HN$^{13}$C/HC$_{3}$N, H$^{13}$CO$^+$/HC$_{3}$N, and HC$^{17}$O$^+$/HC$_{3}$N.
Top panel: dendrogram obtained using Ward’s method. The vertical axis represents the Ward distance, which quantifies the chemical dissimilarity between cores. The horizontal dashed line indicates the adopted linkage-distance cut. The x-axis labels denote the core numbering.
Bottom panels: mean abundance ratios of HN$^{13}$C, H$^{13}$CO$^+$, and HC$^{17}$O$^+$ relative to HC$_3$N for the resulting clusters, plotted as a function of the mean kinetic temperature of each cluster. The clusters correspond to groups of cores with similar abundance ratios.
Error bars represent the standard deviation within each cluster.
}
\label{fig:clusters_cut1}
\end{figure}

\begin{figure}[t]
\centering
\includegraphics[width=\linewidth]{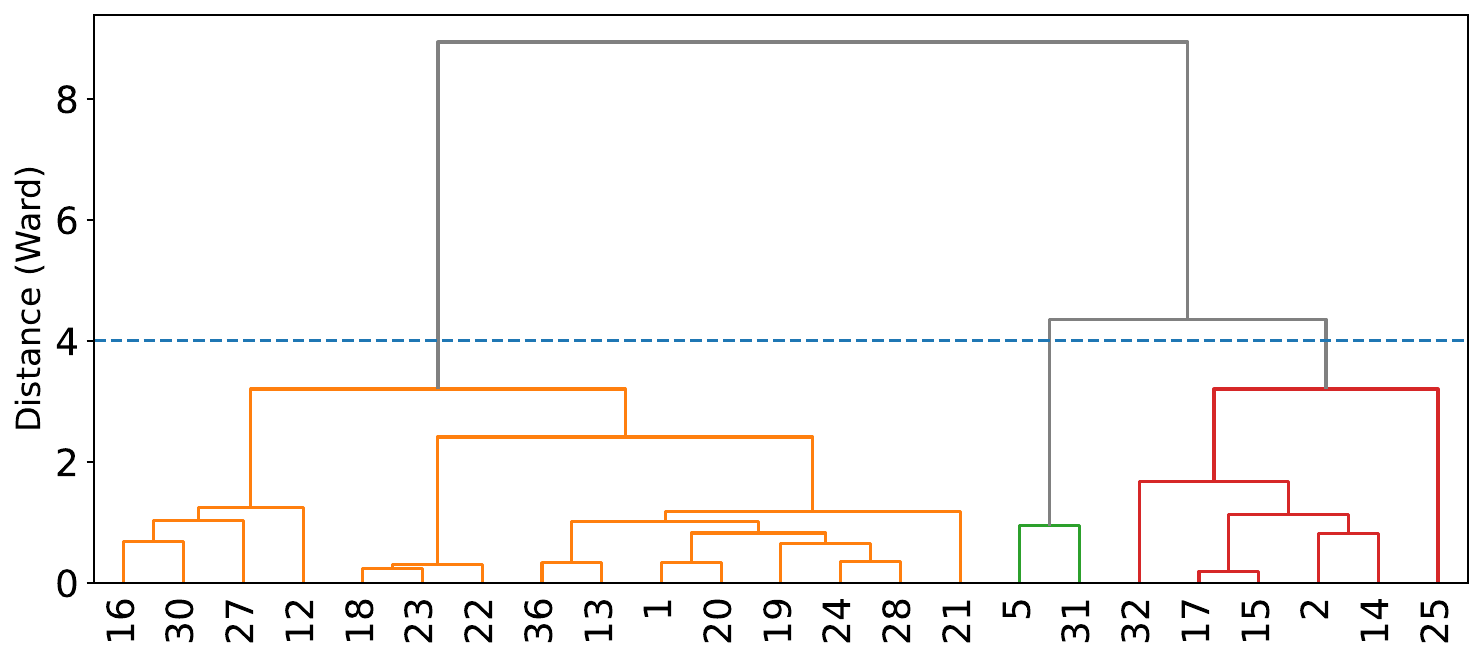}
\vspace{0.2cm}
\includegraphics[width=\linewidth]{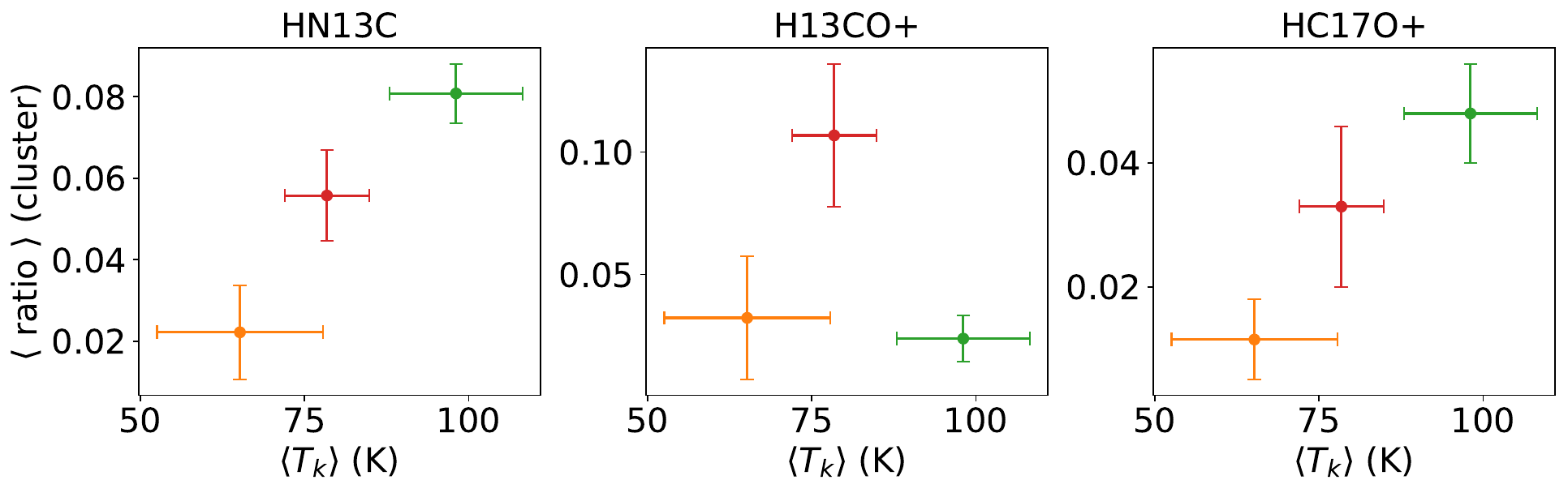}
\caption{
Same as Fig.\,\ref{fig:clusters_cut1}, but adopting a higher linkage-distance cut in the dendrogram, resulting in a smaller number of clusters.
}
\label{fig:clusters_cut2}
\end{figure}

Prior to clustering, all ratios were standardised to a zero mean and unit variance to avoid biases associated with different dynamic ranges. The clustering was carried out using Ward’s method, which groups sources by minimising the intra-cluster variance. In this context, the Ward distance shown in the dendrogram represents the dissimilarity between cores regarding the considered abundance ratios; larger distances indicate more chemically distinct cores. Thus, we are effectively separating the cores based on the chemical conditions reflected by the analysed molecular species. The obtained dendrograms are presented in Figs.\,\ref{fig:clusters_cut1} and \ref{fig:clusters_cut2} (upper panels).
Since the definition of clusters in a hierarchical dendrogram depends on the choice of the linkage-distance cut, which is not unique, we explored two representative cuts of the dendrogram. These cuts are indicated by horizontal dashed lines. It is important to note that the definition of clusters depends on the chosen cut, and therefore the number of groups is not unique.
Figure\,\ref{fig:clusters_cut1} shows the dendrogram with a cut at a low Ward distance, resulting in five clusters of cores and an isolated one, while the cut at a higher linkage-distance (see dendrogram in Fig.\,\ref{fig:clusters_cut2}) yields a smaller number of clusters. The resulting clusters correspond to groups of molecular cores with similar chemical conditions as defined by the used abundance ratios.

To further analyse the clusters in relation to a physical parameter that characterizes them, the bottom panels of Figs.\,\ref{fig:clusters_cut1} and \ref{fig:clusters_cut2} show the average abundance ratios of the clusters as a function of their average temperature. It is worth noting that the kinetic temperature is independent of the clustering method. Therefore, any correlation between the clusters and the average temperature emerges as an independent result of the analysis, rather than being imposed by the clustering method.

\section{Discussion}

The abundances of HN$^{13}$C, HC$^{17}$O$^{+}$, and H$^{13}$CO$^{+}$ clearly increase with the kinetic temperature.
Figure\,\ref{XvsT} shows that such molecular abundances are positively correlated with T$_{\rm k}$. In the case of H$^{13}$CN, although there are only 6 points, the same trend can be observed. The positive correlations between these molecular abundances and the temperature are similar to those found in the case of sulphur-bearing molecules \citep{fontani23,sulfur24}. Such a correlation suggests that the chemistry involved at the production of these simple molecular species should have a similar dependence with the temperature. This dependence may arise through two main mechanisms: first, the thermal desorption of these molecular species or their precursors from dust-grain ice mantles; and second, the activation of gas-phase chemical reactions that effectively produce such species.

It is important to remark that the T$_{\rm k}$ values, which were obtained from our previous study \citep{sulfur24}, are used as a thermal indicator of each core. The cores could present a mixture of thermal conditions, but we are focusing on analysing the molecular species abundances with respect some thermal indicator, and the T$_{\rm k}$ derived from the methanol lines is indeed a good quantity to have an estimation of the average core temperature (see \citealt{martinez26}). More precise calculation of temperature would certainly require observing several transitions of different molecules. However, this becomes impractical due to the large size of the core sample. Therefore, using both methanol lines, which are present throughout the entire sample of cores, is the best compromise achieved to maintain uniformity.
Nevertheless the temperature range obtained for the sample is consistent with early cores embedded in cold molecular clumps; these conditions facilitate a quite rich chemistry, though not as complex as that of hot cores with temperatures above 100 K (e.g., \citealt{chou15,garrod17}).

The HC$_{3}$N abundance does not appear to correlate with temperature. The very low Spearman coefficient ($\rho = 0.09$) with a p-value of 0.65 indicates that there is no statistically significant correlation between X(HC$_{3}$N) and the temperature in the analysed cores sample.
For instance, these results are in agreement with \citet{taniguchi18}, who found a slight decrease in X(HC$_{3}$N) from high-mass starless cores to high-mass protostellar objects using single-dish observations of the J=5–4 line, suggesting that X(HC$_{3}$N) may be relatively insensitive to temperature changes.
Although the $\ce{HC3N}$ transition possesses a high excitation energy (E$_{\rm u} = 306.9$ K), possibly indicating non-thermalised emission (further analysed in Sect.\,\ref{Sectradex}), its fractional abundance remains relatively flat. Chemically, this plateau in the $\rm 40-100~K$ range reflects a steady state between gas-phase production and grain-surface depletion. At  T $\sim$ 50~K, acetylene sublimation boosts $\ce{HC3N}$ production via the main neutral-neutral reaction pathway: $\ce{C2H2 + CN}$ $\rightarrow$~ $\ce{HC3N + H}$  \citep{wang23}. However, this efficient synthesis is offset by simultaneous freeze-out, which enriches the icy grain mantles \citep{taniguchi19} and thus probably keeps the observed abundance relatively constant. Because our early-stage cores lack the $\rm T>100~K$ threshold required to thermally desorb this solid reservoir, probably the ultimate $\ce{HC3N}$ abundance peak is not yet reached.


All the comparisons among abundances shown in Fig.\,\ref{fig:abundancias} present a positive correlation; in some of them, the increase in X vs. X correlates with the
kinetic temperature. These are the cases of X(H$^{13}$CO$^{+}$) vs X(HC$^{17}$O$^{+}$), X(H$^{13}$CO$^{+}$) vs X(HN$^{13}$C), X(HN$^{13}$C) vs X(HC$^{17}$O$^{+}$), and although there are few points, it also happens in X(H$^{13}$CN) vs X(HC$^{17}$O$^{+}$), and X(H$^{13}$CO$^{+}$) vs X(H$^{13}$CN).  
This phenomenon shows that as the temperature increases, the abundances also increase, which supports,
as mentioned above, the hypothesis that the production of these molecular species should have a similar dependence with the temperature (in a range of 40--100 K) in early star-forming molecular cores. Interestingly, even though a weak positive correlation might be observed between the X($\ce{HC3N}$) and X($\ce{HN^13C}$) or X($\ce{H^13CN}$), this behaviour remains consistent with their respective gas-phase chemistries. Chemical pathways linking these species do exist (e.g., $\ce{C2H + HNC} \rightarrow \ce{HC3N + H}$, \citealt{taniguchi16}), but they are considered secondary mechanisms of production in these environments. Because simple cyanides are not appreciably depleted to form the bulk of $\ce{HC3N}$, there is no strong chemical competition between them. Instead, both networks are enhanced by the thermal desorption, allowing their abundances to co-evolve without mutual destruction.

Additionally, the aforementioned positive correlations between abundances, as well as between abundances and temperature, are independently verified via the Spearman matrix method illustrated in Fig.\,\ref{fig:corr_matrix} for all analysed molecules excepting the H$^{13}$CN.

The hierarchical clustering analysis provides an independent perspective to interpret the chemical behaviour of the cores based on the analysed molecular species. 
When the sources are grouped according to their abundance ratios, the resulting clusters show differences in their kinetic temperatures. 
In particular, the clusters associated with lower temperatures exhibit lower mean abundance ratios of HN$^{13}$C/HC$_3$N and HC$^{17}$O$^+$/HC$_3$N, whereas clusters with higher kinetic temperatures show larger average ratios. This trend indicates that the chemical differentiation identified by the clustering analysis is closely linked to the thermal properties of the cores. It is worth noting that despite the differences in the number of clusters, both cuts in the dendrogram show consistent behaviour. 
These results support, through an independent statistical approach based on molecular abundance ratios, the correlations found in Sect.\,\ref{results}.

From the positive correlations found between the HN$^{13}$C, H$^{13}$CO$^+$, and HC$^{17}$O$^+$ abundances and temperature, we conclude that the production and survival of such molecular species may become progressively more efficient as the gas temperature rises in these early star-forming environments. Specifically, $\ce{CO}$ sublimation ($\rm T \gtrsim 30~K$) injects this primary precursor into the gas phase, directly increasing formylium isotopologues through rapid ion-molecule reactions ($\ce{H3+ + CO} \rightarrow \ce{HCO+ + H2}$, \citealt{jor20}). Concurrently, the thermal release of nitrogen-bearing reactants sustains the parallel production of simple cyanides in the warm envelope. In agreement with time-dependent chemical models \citep{wang25}, the strictly positive correlations observed among these species confirm a lack of destructive chemical competition in these cores.


Regarding the H$^{13}$CN and HN$^{13}$C ratios presented in Fig.\,\ref{isomers}, it is important to mention that the limited data points prevent us from drawing statistically robust conclusions.
However, we can note that from the six analysed cores, no correlation is observed between the ratios and temperature, contrary to other studies \citep{hacar20,pazu22,nai23,nai24}. Nevertheless, those studies show correlations for temperature regimes below 40 K. Additionally, we observe that the H$^{13}$CN/HN$^{13}$C abundance ratios are greater than unity, which is consistent with regions of increasing temperature where HNC is preferentially destroyed over HCN \citep{shcilke92,hirota98}.


\subsection{Remarks on HC$^{17}$O$^{+}$} 

It is important to highlight that very few works in the literature report observations of HC$^{17}$O$^+$. Here, we present a valuable sample of cores with positive detection of such scarce molecular species. 
Among the HCO$^{+}$ isotopologues, the HC$^{17}$O$^+$ seems to be the most difficult to detect \citep{caselli02,ferrer22,mendoza23}. For instance, \citet{mendoza23} detect H$^{17}$CO$^+$ (single dish observations) toward the hot molecular core G331.512-0.103, obtaining a column density of approximately 1.5 $\times$ 10$^{12}$ cm$^{-2}$, a similar value to those of many of our studied cores. 

Across the core sample, the abundance ratio H$^{13}$CO$^{+}$/HC$^{17}$O$^{+}$ averages 3.4, ranging from 0.3 to 8.8. This ratio can be used to study oxygen isotopic fractionation in the ISM. We can derive a $^{16}$O/$^{17}$O ratio by assuming a standard $^{12}$C/$^{13}$C value. Considering the galactocentric distance of the cores (obtained from their distances and galactic coordinates), a $^{12}$C/$^{13}$C ratio of 50 is a reasonable value, consistent with the standard relations presented in the literature \citep{wilson94,milam05}. Using this value, we obtain an average $^{16}$O/$^{17}$O ratio of 170, with a minimum of 16 and a maximum of 440 across our sample of cores. These values are significantly lower than the expected Galactic range ($\sim$1000--2000) predicted and/or measured in some works \citep{wilson94,wout08,persson2018}. This result suggests that the H$^{13}$CO$^{+}$ line must be optically thick in these regions, contrary to common assumptions. Indeed, as \citet{bizzo24} point out, the H$^{13}$CO$^{+}$ can be optically thick in numerous sources. 

It is important to note that a detailed visual inspection of the H$^{13}$CO$^+$ spectra in all sources reveals no obvious signs of high optical depth, such as flat-topped profiles or self-absorption features. However, we note that sources with the lowest H$^{13}$CO$^+$/HC$^{17}$O$^+$ ratios (e.g., core\,6) have spectra more densely populated with molecular lines than sources with the highest ratios (e.g., core\,12). This finding indirectly supports the hypothesis of high opacity in the H$^{13}$CO$^+$ line in the densest and most chemically active regions.

\subsection{Non-LTE analysis for the HC$_{3}$N}
\label{Sectradex}

We carried out a non-LTE study for HC$_{3}$N J=37--36 to obtain its column density through the RADEX code \citep{van07}. The used inputs are: the line peak intensity (T$_{\rm p}$) and the FWHM line width ($\Delta$v) measured from the spectra in this work, and the kinetic temperatures (T$_{\rm k}$) and the volume density of the regions (n$_{\rm H_2}$) both obtained in our previous work \citep{sulfur24}.
T$_{\rm p}$ and  $\Delta$v were obtained from Gaussian fits to the observed lines. The T$_{\rm p}$, in K, was obtained from the measured peak intensity I$_{\rm peak}$, in Jy beam$^{-1}$, using the equation $\rm T_{p} = 1.222 \times 10^{3} \frac{I_{peak}}{\upnu^{2} \uptheta_{maj} \uptheta_{min}}$, where $\upnu$ is the line rest frequency, and  $\rm \uptheta_{maj}$ and $\rm \uptheta_{min}$ are the half-power beam widths along the major and minor axes, respectively.

\begin{figure}[h]
\centering
\includegraphics[width=8.5cm]{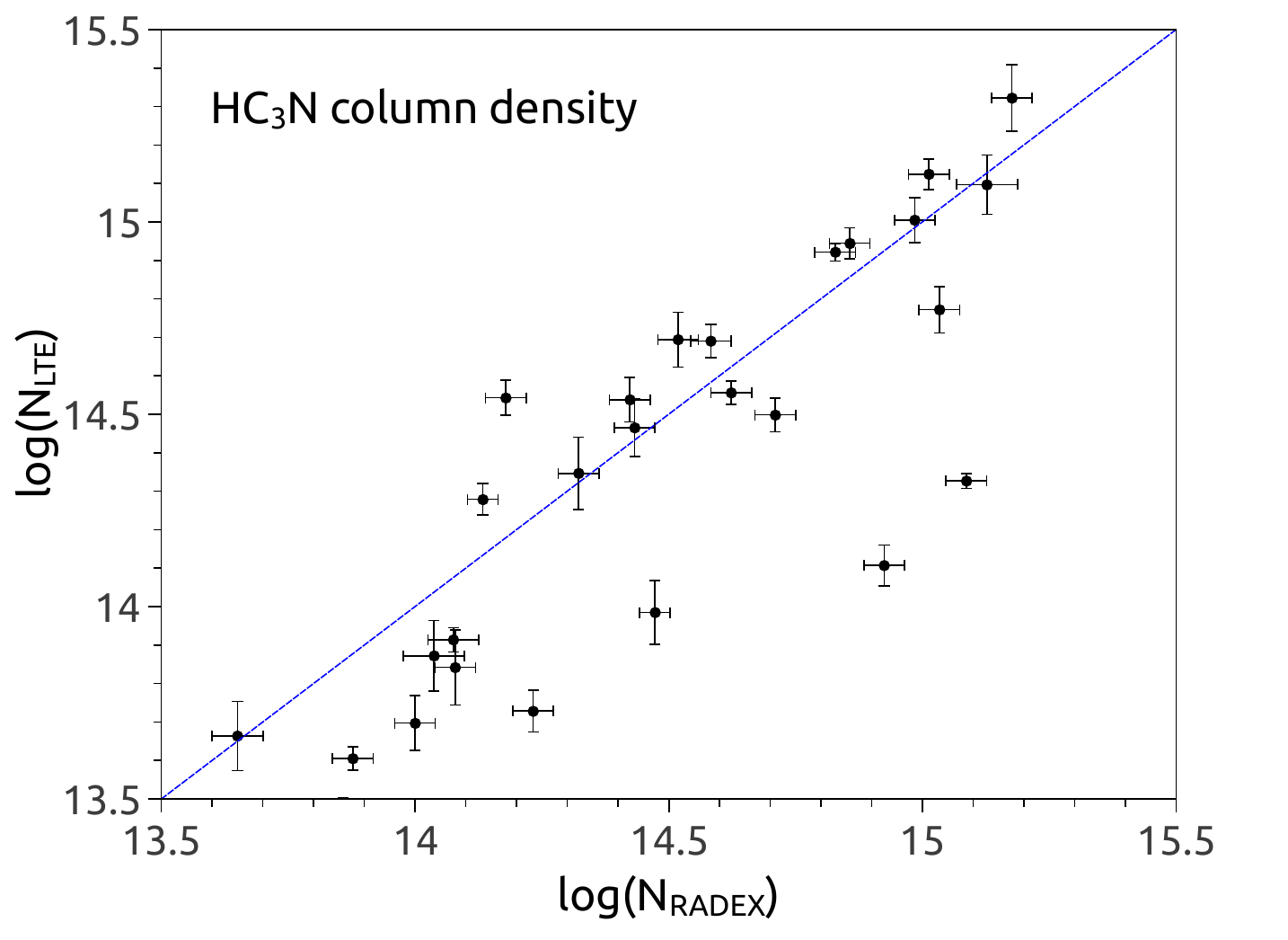}
\caption{Comparisons between the column densities obtained for the HC$_{3}$N J=37--36 line: log(N$_{\rm LTE}$) vs. log(N$_{\rm RADEX}$).  The unity line is included (dashed blue).  }
\label{Nradexlte}
\end{figure}

Figure\,\ref{Nradexlte} shows the obtained N$_{\rm RADEX}$ values vs N$_{\rm LTE}$ in a log-log diagram. For comparison, the unity line (dashed blue) is also included. Although some points lie slightly off the unit line, the observed column densities derived from both methods are in quite good agreement. The points furthest from the unity line show N$_{\text{RADEX}}$ values higher than N$_{\text{LTE}}$, suggesting that in those cases the HC$_3$N line might not be fully thermalized. However, using the results obtained from RADEX alters neither the X(HC$_3$N) abundance correlations nor its lack of correlation with temperature. Consequently, both the Spearman trends and the resulting dendrogram and cluster distributions from Sect.\,\ref{stats} remain essentially unchanged.
 
\section{Summary}

Given that understanding the chemistry in the early stages of star formation is a matter of importance we analysed the chemical behaviour of simple cyanides and formylium ions isotopologues in a sample of 37 massive infrared-quiet molecular cores. From ALMA observations at Band 7, we calculated abundances of HC$_3$N, H$^{13}$CN, HN$^{13}$C, H$^{13}$CO$^+$, and HC$^{17}$O$^+$, which are useful to study the early chemistry in molecular cores. Our main results can be summarised as follows:

(i) We found that the abundances of HN$^{13}$C, H$^{13}$CO$^+$, and HC$^{17}$O$^+$ show positive correlations with the kinetic temperature, showing that the production and survival of such molecular species may become progressively more efficient as the gas temperature rises in these early star-forming environments. In the case of H$^{13}$CN the same is suggested, but we have too few data points to be conclusive. The temperature-driven chemical regulation may be a common characteristic of simple molecular species in young massive cores.

(ii) The analysis of the H$^{13}$CN/HN$^{13}$C ratios in the six cores where H$^{13}$CN was measured does not suggest any correlation with temperature in the 40 to 100 K range, as it is found in other works for lower temperatures.

(iii) The abundance of HC$_3$N does not exhibit a temperature dependence. This is compatible with a chemical steady state between gas-phase production and grain-surface depletion in the studied temperature range for this molecular species. 

(iv) We presented a hierarchical clustering method based on abundance ratios to study chemical differentiation among the cores. We found different groups of cores based on their HN$^{13}$C/HC$_{3}$N, H$^{13}$CO$^+$/HC$_{3}$N, and HC$^{17}$O$^+$/HC$_{3}$N abundance ratios, which in addition correlate with the temperature. This represents a novel method for chemically characterizing large samples of interstellar structures.

(v) We detected HC$^{17}$O$^+$ emission in 28 cores from our sample; given the scarce reports of this species in the literature, this represents a valuable sample of regions for its study. Estimates of $^{16}$O/$^{17}$O from the H$^{13}$CO$^+$/HC$^{17}$O$^+$ ratios suggest that the H$^{13}$CO$^+$ line may have high optical depths, contrary to common assumptions in the literature. 


\begin{acknowledgements}

We thank the anonymous referee for their useful comments. R.D.T. and N.C.M. are doctoral fellows of CONICET. S.P. and M.E.O. are members of the {\it Carrera del Investigador Cient\'\i fico} of CONICET, Argentina.  This work was partially supported by the Argentina grants PIP 2021 11220200100012 awarded by CONICET. 
This paper makes use of the following ALMA data: ADS/JAO.ALMA$\#$2017.1.00914.S ALMA is a partnership of ESO (representing its member states), NSF (USA) and NINS (Japan), together with NRC (Canada), MOST and ASIAA (Taiwan), and KASI (Republic of Korea), in cooperation with the Republic of Chile. The Joint ALMA Observatory is operated by ESO, AUI/NRAO and NAOJ. The National Radio Astronomy Observatory is a facility of the National Science Foundation operated under cooperative agreement by Associated Universities, Inc.

\end{acknowledgements}

%
%


\bibliographystyle{aa}  
\bibliography{ref}
\IfFileExists{\jobname.bbl}{}
{\typeout{}
\typeout{****************************************************}
\typeout{****************************************************}
\typeout{** Please run "bibtex \jobname" to optain}
\typeout{** the bibliography and then re-run LaTeX}
\typeout{** twice to fix the references!}
\typeout{****************************************************}
\typeout{****************************************************}
\typeout{}
}
\label{lastpage}

\begin{appendix} 

\section{Spectral line profiles}
\label{spectraexample}

\begin{figure}[h!]
\centering

\includegraphics[width=1\linewidth]{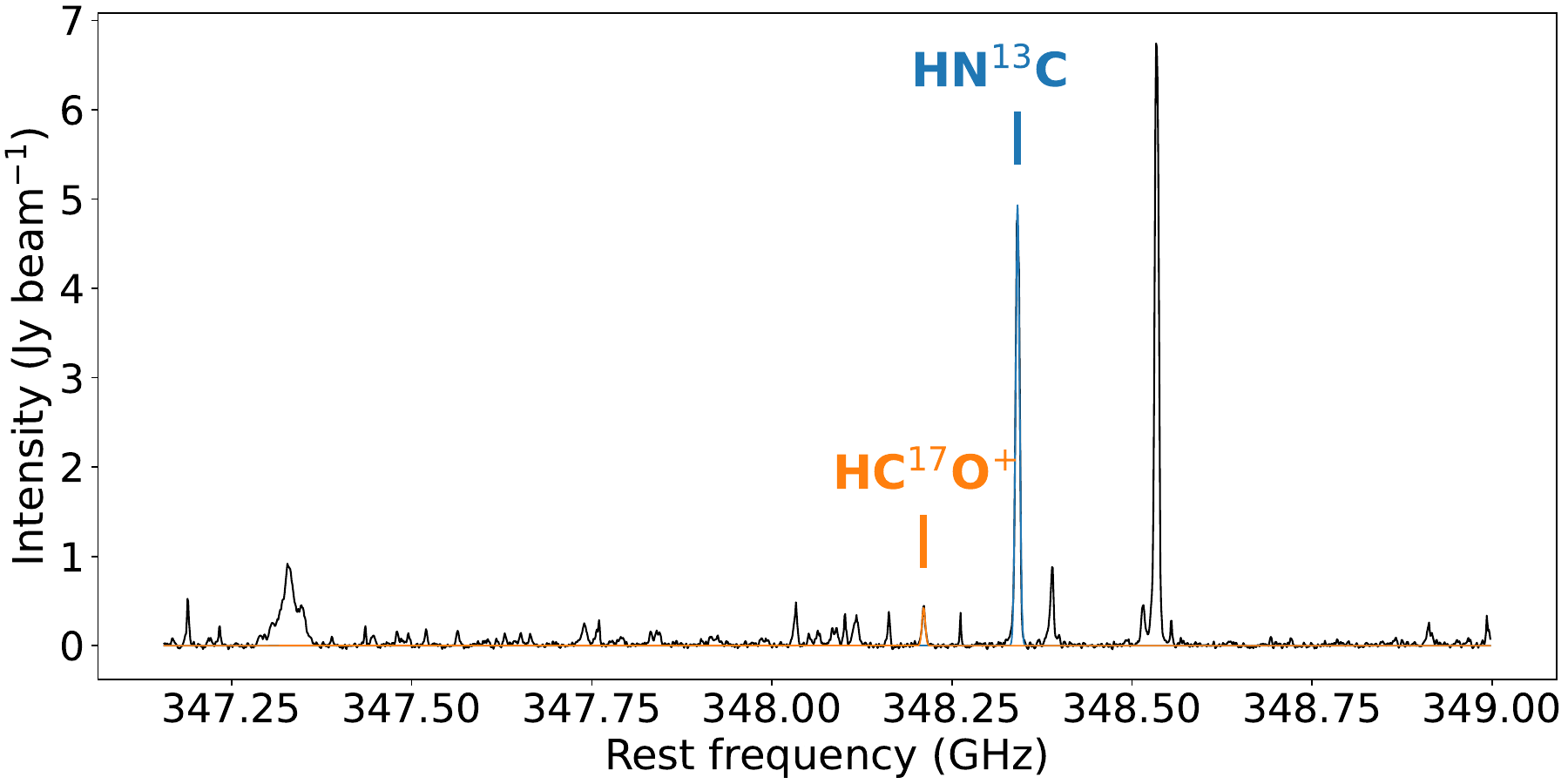}
\includegraphics[width=1\linewidth]{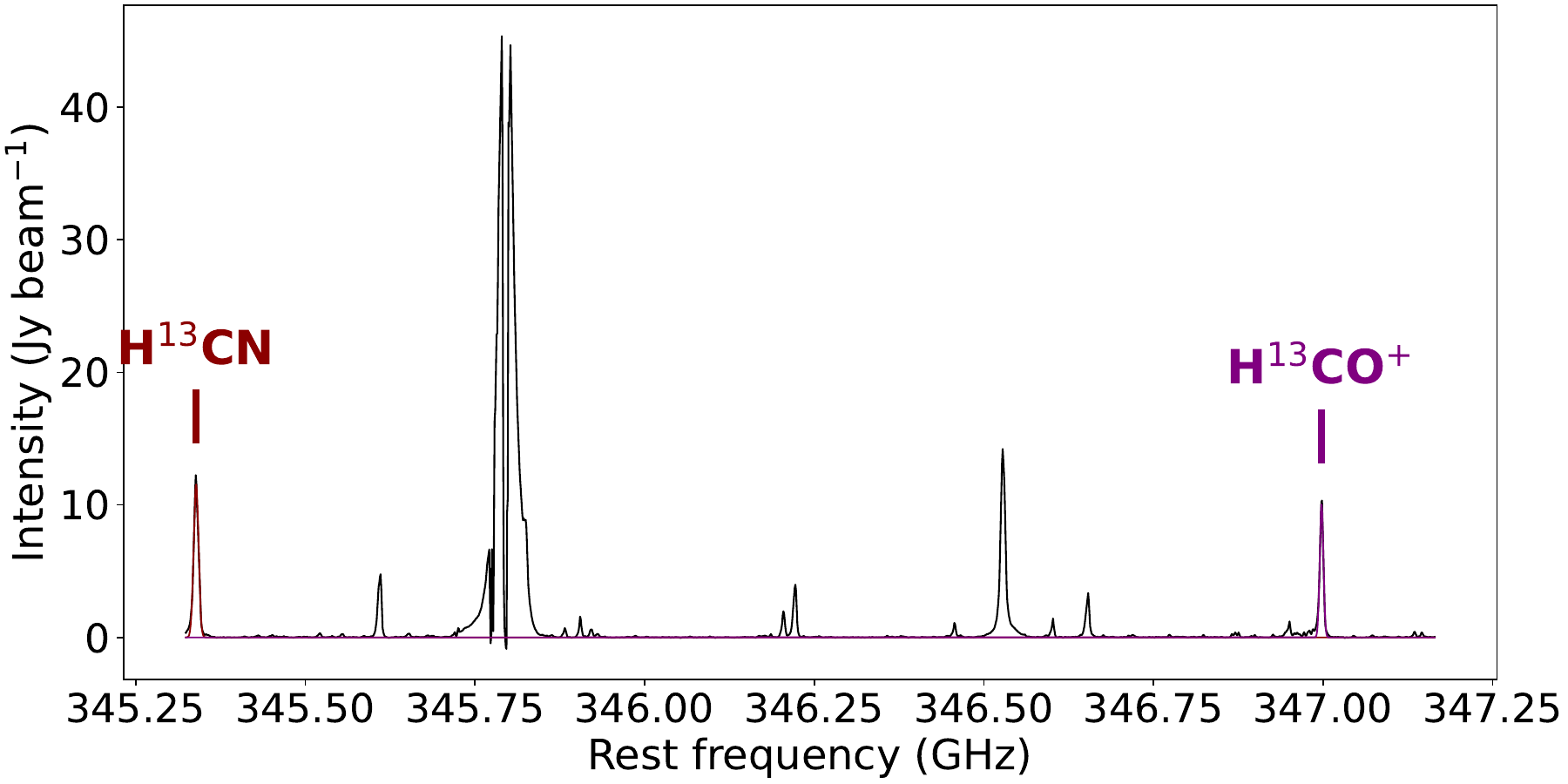}


\includegraphics[width=1\linewidth]{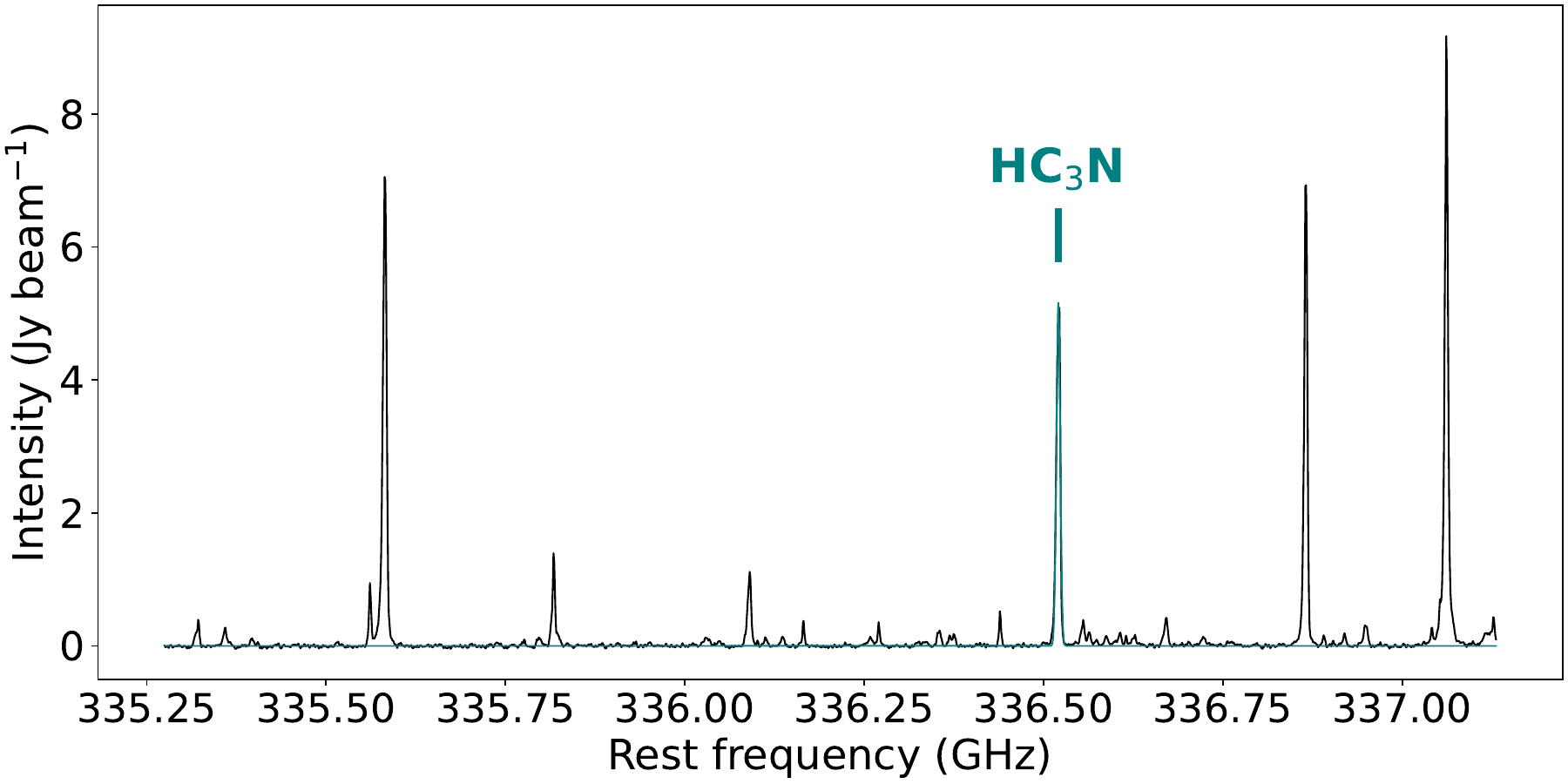}

\caption{
Spectra obtained from the region G310.1440+0.7592 (core\,22). Molecular transitions of
HC$_3$N, H$^{13}$CN, HN$^{13}$C, H$^{13}$CO$^{+}$, 
and HC$^{17}$O$^{+}$ are labelled. Gaussian fits to the selected lines 
are shown in colours.
}
\label{fig:spectra_G310.1440+0.7592}
\end{figure}

Figure\,\ref{fig:spectra_G310.1440+0.7592} displays as an illustrative example the spectra
obtained towards G310.1440$+$0.7592 (core\,22). The analysed lines are indicated and the Gaussian fittings are included.

\section{Molecular lines parameters}
\label{append0}

Tables\,\ref{tab:hc3n}, \ref{tab:h13cN-hc13N}, and \ref{tab:h13co-hc17o} present the parameters obtained from Gaussian fits to the molecular lines in all cores.

\begin{table}[h!]
\centering
\small
\setlength{\tabcolsep}{2pt}
\caption{Line parameters obtained from Gaussian fits to the HC$_3$N J=37--36.}
\label{tab:hc3n}
\begin{tabular}{lcccccc}
\hline
\hline
Core & Peak & err. & $\Delta$ v & err. & W & err.  \\

  & (Jy beam$^{-1}$) &  & (km s$^{-1}$) &  & (Jy beam$^{-1}$ km s$^{-1}$) & \\
\hline
1  & -    & -    & -    & -    & -     & -    \\
2  & 1.96 & 0.07 & 6.71 & 0.26 & 13.98 & 0.54 \\
3  & 1.00 & 0.04 & 6.63 & 0.34 & 7.07  & 0.36 \\
4  & 1.71 & 0.03 & 11.56& 0.28 & 21.07 & 0.50 \\
5  & -    & -    & -    & -    & -     & -    \\
6  & 2.05 & 0.04 & 11.14& 0.28 & 24.25 & 0.60 \\
7  & 4.57 & 0.09 & 6.74 & 0.15 & 32.74 & 0.73 \\
8  & 0.27 & 0.01 & 7.09 & 0.37 & 2.00  & 0.10 \\
9  & 1.37 & 0.04 & 5.43 & 0.19 & 7.91  & 0.28 \\
10 & -    & -    & -    & -    & -     & -    \\
11 & 1.76 & 0.05 & 6.55 & 0.20 & 12.22 & 0.37 \\
12 & -    & -    & -    & -    & -     & -    \\
13 & 0.58 & 0.01 & 7.03 & 0.16 & 4.31  & 0.10 \\
14 & 2.81 & 0.04 & 11.42& 0.22 & 34.18 & 0.65 \\
15 & 0.31 & 0.01 & 10.02& 0.44 & 3.33  & 0.15 \\
16 & 0.31 & 0.01 & 9.32 & 0.54 & 3.06  & 0.18 \\
17 & 0.94 & 0.02 & 8.16 & 0.25 & 8.16  & 0.25 \\
18 & 0.92  & 0.03    & 4.36    & 0.15   &  4.28  & 0.15   \\
19 & 0.40 & 0.02 & 10.04& 0.49 & 4.30  & 0.21 \\
20 & 5.38 & 0.11 & 9.44 & 0.23 & 53.98 & 1.32 \\
21 & 0.70 & 0.01 & 8.95 & 0.23 & 6.66  & 0.17 \\
22 & 5.17 & 0.18 & 5.59 & 0.23 & 30.73 & 1.26 \\
23 & 2.04 & 0.03 & 10.90& 0.20 & 23.70 & 0.44 \\
24 & 0.26 & 0.01 & 6.75 & 0.44 & 1.85  & 0.12 \\
25 & 0.10 & 0.01 & 6.80 & 0.69 & 0.74  & 0.07 \\
26 & 0.26 & 0.01 & 5.87 & 0.34 & 1.64  & 0.09 \\
27 & -    & -    & -    & -    & -     & -    \\
28 & 0.56 & 0.01 & 8.32 & 0.22 & 4.97  & 0.13 \\
29 & 10.32& 0.27 & 9.14 & 0.29 & 100.24& 3.18 \\
30 & 20.99& 0.49 & 8.55 & 0.24 & 190.83& 5.35 \\
31 & 1.25 & 0.04 & 6.77 & 0.28 & 9.03  & 0.37 \\
32 & 4.64 & 0.16 & 8.38 & 0.33 & 41.30 & 1.63 \\
33 & 0.43 & 0.01 & 11.74& 0.41 & 5.42  & 0.19 \\
34 & -    & -    & -    & -    & -     & -    \\
35 & -    & -    & -    & -    & -     & -    \\
36 & -    & -    & -    & -    & -     & -    \\
37 & 3.62 & 0.11 & 6.56 & 0.24 & 25.21 & 0.91 \\
\hline
\multicolumn{6}{l}{A `-' means that no emission above the noise level was detected. } \\
\end{tabular}
\end{table}

\begin{table*}[h!]
\centering
\setlength{\tabcolsep}{2.5pt}
\caption{Line parameters obtained from Gaussian fits to the H$^{13}$CN $\rm (v=0)$ and HN$^{13}$C J=4--3 lines.}
\label{tab:h13cN-hc13N}
\begin{tabular}{lcccccc|cccccc}
\hline
 & \multicolumn{6}{c}{H$^{13}$CN (*)} 
 & \multicolumn{6}{c}{HN$^{13}$C} \\
\hline
Core 
& Peak & err. & $\Delta$v & err. & W & err.
& Peak & err. & $\Delta$v & err. & W & err. \\
 & (Jy beam$^{-1}$) &  & (km s$^{-1}$) &  & (Jy beam$^{-1}$ km s$^{-1}$) & 
 & (Jy beam$^{-1}$) &  & (km s$^{-1}$) &  & (Jy beam$^{-1}$ km s$^{-1}$) &  \\
\hline
1  & 2.38 & 0.10 & 8.26 & 0.39 & 20.72 & 0.99 & 0.59 & 0.02 & 4.58 & 0.17 & 2.90 & 0.11 \\
2  & -    & -    & -    & -    & -     & -    & 2.79 & 0.19 & 4.35 & 0.34 & 12.90 & 1.00 \\
3  & -    & -    & -    & -    & -     & -    & 1.40 & 0.05 & 4.39 & 0.18 & 6.52  & 0.26 \\
4  & -    & -    & -    & -    & -     & -    & 1.57 & 0.03 & 8.66 & 0.21 & 14.46 & 0.35 \\
5  & -    & -    & -    & -    & -     & -    & 0.46 & 0.17 & 1.05 & 0.41 & 0.51  & 0.20 \\
6  & -    & -    & -    & -    & -     & -    & 1.35 & 0.08 & 10.76& 0.81 & 15.46 & 1.16 \\
7  & -    & -    & -    & -    & -     & -    & 3.00 & 0.10 & 7.24 & 0.28 & 23.09 & 0.91 \\
8  & -    & -    & -    & -    & -     & -    & 0.23 & 0.02 & 4.61 & 0.53 & 1.12  & 0.13 \\
9  & -    & -    & -    & -    & -     & -    & 0.86 & 0.04 & 4.64 & 0.27 & 4.25  & 0.24 \\
10 & -    & -    & -    & -    & -     & -    & 0.51 & 0.48 & 0.93 & 0.07 & 0.51  & 0.42 \\
11 & -    & -    & -    & -    & -     & -    & 1.29 & 0.03 & 5.52 & 0.17 & 7.59  & 0.23 \\
12 & -    & -    & -    & -    & -     & -    & 0.54 & 0.03 & 2.79 & 0.17 & 1.61  & 0.10 \\
13 & -    & -    & -    & -    & -     & -    & 0.70 & 0.03 & 2.79 & 0.14 & 2.08  & 0.10 \\
14 & -    & -    & -    & -    & -     & -    & 2.32 & 0.12 & 8.81 & 0.56 & 21.73 & 1.37 \\
15 & -    & -    & -    & -    & -     & -    & 0.36 & 0.02 & 4.98 & 0.32 & 1.92  & 0.12 \\
16 & -    & -    & -    & -    & -     & -    & 0.67 & 0.02 & 4.92 & 0.21 & 3.50  & 0.15 \\
17 & -    & -    & -    & -    & -     & -    & 0.53 & 0.08 & 3.35 & 0.57 & 1.90  & 0.32 \\
18 & -    & -    & -    & -    & -     & -    & 1.14 & 0.03 & 4.95 & 0.16 & 6.02  & 0.19 \\
19 & -    & -    & -    & -    & -     & -    & 0.36 & 0.02 & 5.65 & 0.42 & 2.14  & 0.16 \\
20 & -    & -    & -    & -    & -     & -    & 2.58 & 0.03 & 11.53& 0.15 & 31.68 & 0.42 \\
21 & -    & -    & -    & -    & -     & -    & 0.75 & 0.03 & 7.13 & 0.34 & 5.66  & 0.27 \\
22 & 11.56& 0.21 & 7.28 & 0.15 & 88.81 & 1.83 & 4.94 & 0.10 & 5.66 & 0.13 & 29.71 & 0.68 \\
23 & -    & -    & -    & -    & -     & -    & 0.92 & 0.07 & 13.52& 1.78 & 13.28 & 1.75 \\
24 & 2.48 & 0.16 & 4.27 & 0.32 & 11.18 & 0.84 & 1.33 & 0.04 & 1.62 & 0.05 & 2.29  & 0.07 \\
25 & -    & -    & -    & -    & -     & -    & 2.04 & 0.03 & 1.65 & 0.03 & 3.57  & 0.07 \\
26 & -    & -    & -    & -    & -     & -    & 0.42 & 0.02 & 5.03 & 0.34 & 2.26  & 0.15 \\
27 & 0.80 & 0.04 & 2.70 & 0.06 & 6.81  & 0.14 & 0.93 & 0.03 & 2.36 & 0.10 & 2.34  & 0.10 \\
28 & -    & -    & -    & -    & -     & -    & 0.53 & 0.04 & 4.44 & 0.41 & 2.48  & 0.23 \\
29 & -    & -    & -    & -    & -     & -    & 4.98 & 0.12 & 10.33& 0.32 & 54.66 & 1.71 \\
30 & -    & -    & -    & -    & -     & -    & 15.78& 0.54 & 8.91 & 0.38 & 149.58& 6.31 \\
31 & 5.21 & 0.21 & 7.03 & 0.32 & 38.66 & 1.78 & 1.43 & 0.10 & 5.07 & 0.41 & 7.74  & 0.63 \\
32 & -    & -    & -    & -    & -     & -    & 3.16 & 0.11 & 10.29& 0.46 & 34.53 & 1.54 \\
33 & -    & -    & -    & -    & -     & -    & 0.47 & 0.03 & 8.05 & 0.64 & 4.06  & 0.32 \\
34 & -    & -    & -    & -    & -     & -    & 0.12 & 0.01 & 1.96 & 0.28 & 0.24  & 0.04 \\
35 & -    & -    & -    & -    & -     & -    & 0.35 & 0.01 & 1.37 & 0.05 & 0.52  & 0.02 \\
36 & -    & -    & -    & -    & -     & -    & 0.19 & 0.02 & 1.67 & 0.19 & 0.34  & 0.04 \\
37 & 8.44 & 0.29 & 7.51 & 0.30 & 66.76 & 2.68 & 3.56 & 0.12 & 6.48 & 0.25 & 24.54 & 0.96 \\
\hline 
\multicolumn{12}{l}{(*) The lack of H$^{13}$CN parameters in many cores is because in such cases, the spectral} \\
\multicolumn{12}{l}{window where this line lies is truncated near 345.3 GHz, leaving out its emission.} 
\end{tabular}
\end{table*}

\begin{table*}[h!]
\centering
\setlength{\tabcolsep}{2.5pt}
\caption{Line parameters obtained from Gaussian fits to the H$^{13}$CO$^{+}$ and HC$^{17}$O$^{+}$ J=4--3 lines.}
\label{tab:h13co-hc17o}
\begin{tabular}{lcccccc|cccccc}
\hline
 & \multicolumn{6}{c}{H$^{13}$CO$^{+}$} 
 & \multicolumn{6}{c}{HC$^{17}$O$^{+}$} \\
\hline
Core 
& Peak & err. & $\Delta$v & err. & W & err.
& Peak & err. & $\Delta$v & err. & W & err. \\
 & (Jy beam$^{-1}$) &  & (km s$^{-1}$) &  & (Jy beam$^{-1}$ km s$^{-1}$) & 
 & (Jy beam$^{-1}$) &  & (km s$^{-1}$) &  & (Jy beam$^{-1}$ km s$^{-1}$) &  \\
\hline
1  & 4.10 & 0.04 & 3.31 & 0.04 & 14.39 & 0.16 & 0.17 & 0.02 & 3.03 & 0.32 & 0.55 & 0.06 \\
2  & 12.41& 0.45 & 2.76 & 0.12 & 36.25 & 1.52 & 0.69 & 0.03 & 2.90 & 0.16 & 2.12 & 0.12 \\
3  & 6.52 & 0.13 & 3.80 & 0.09 & 26.28 & 0.62 & 0.30 & 0.02 & 4.12 & 0.35 & 1.32 & 0.11 \\
4  & 3.57 & 0.11 & 3.29 & 0.12 & 12.44 & 0.46 & -    & -    & -    & -    & -    & -    \\
5  & 7.51 & 0.30 & 1.42 & 0.07 & 11.29 & 0.53 & 0.22 & 0.02 & 1.76 & 0.23 & 0.40 & 0.05 \\
6  & 2.28 & 0.18 & 2.62 & 0.24 & 6.32 & 0.57 & 0.43 & 0.01 & 7.40 & 0.23 & 3.36 & 0.11 \\
7  & 5.11 & 0.31 & 4.64 & 0.32 & 25.16 & 1.76 & - & - & - & - & - & - \\
8  & 2.34 & 0.12 & 2.87 & 0.17 & 7.10 & 0.42 & - & - & - & - & - & - \\
9  & 7.30 & 0.12 & 3.58 & 0.07 & 27.65 & 0.52 & - & - & - & - & - & - \\
10 & 7.16 & 0.29 & 1.43 & 0.07 & 10.84 & 0.51 & 0.22 & 0.03 & 1.73 & 0.24 & 0.40 & 0.06 \\
11 & 4.94 & 0.12 & 4.71 & 0.13 & 24.66 & 0.68 & - & - & - & - & - & - \\
12 & 4.73 & 0.07 & 3.33 & 0.06 & 16.71 & 0.30 & 0.11 & 0.02 & 2.79 & 0.50 & 0.32 & 0.06 \\
13 & 5.94 & 0.09 & 3.49 & 0.06 & 21.95 & 0.37 & 0.19 & 0.01 & 3.72 & 0.28 & 0.76 & 0.06 \\
14 & 6.55 & 0.22 & 5.25 & 0.20 & 36.44 & 1.42 & 0.63 & 0.03 & 6.05 & 0.30 & 4.04 & 0.20 \\
15 & 3.41 & 0.05 & 3.45 & 0.06 & 12.46 & 0.22 & 0.10 & 0.01 & 4.67 & 0.53 & 0.52 & 0.06 \\
16 & 3.18 & 0.06 & 3.71 & 0.08 & 12.51 & 0.28 & 0.09 & 0.01 & 5.00 & 0.83 & 0.50 & 0.08 \\
17 & 5.33 & 0.08 & 2.82 & 0.05 & 15.92 & 0.28 & 0.23 & 0.01 & 3.89 & 0.24 & 0.94 & 0.06 \\
18 & 4.42 & 0.10 & 4.98 & 0.13 & 23.30 & 0.62 & 0.21 & 0.01 & 3.71 & 0.25 & 0.81 & 0.05 \\
19 & 4.35 & 0.06 & 4.47 & 0.07 & 20.60 & 0.30 & 0.16 & 0.01 & 4.19 & 0.36 & 0.72 & 0.06 \\
20 & 4.18 & 0.23 & 7.09 & 0.46 & 31.40 & 2.03 & 0.76 & 0.04 & 8.33 & 0.53 & 6.71 & 0.42 \\
21 & 2.60 & 0.06 & 4.26 & 0.11 & 11.73 & 0.31 & 0.15 & 0.01 & 5.11 & 0.48 & 0.82 & 0.08 \\
22 & 10.09& 0.24 & 5.40 & 0.15 & 57.76 & 1.59 & 0.42 & 0.01 & 4.66 & 0.17 & 2.10 & 0.08 \\
23 & 5.25 & 0.16 & 7.07 & 0.25 & 39.36 & 1.37 & 0.69 & 0.02 & 4.97 & 0.15 & 3.67 & 0.11 \\
24 & 3.18 & 0.07 & 2.48 & 0.06 & 8.34 & 0.20 & 0.10 & 0.01 & 4.11 & 0.67 & 0.42 & 0.07 \\
25 & 1.33 & 0.06 & 2.29 & 0.11 & 3.22 & 0.15 & 0.30 & 0.02 & 1.71 & 0.12 & 0.54 & 0.04 \\
26 & 2.93 & 0.06 & 3.24 & 0.08 & 10.06 & 0.24 & 0.09 & 0.01 & 6.47 & 1.15 & 0.62 & 0.11 \\
27 & 2.39 & 0.04 & 2.70 & 0.06 & 6.81 & 0.14 & 0.13 & 0.03 & 1.47 & 0.42 & 0.20 & 0.06 \\
28 & 4.50 & 0.16 & 3.38 & 0.13 & 16.11 & 0.64 & 0.14 & 0.01 & 2.65 & 0.32 & 0.39 & 0.05 \\
29 & 3.24 & 0.12 & 7.08 & 0.31 & 24.29 & 1.07 & 0.78 & 0.02 & 6.10 & 0.20 & 5.04 & 0.16 \\
30 & -    & -    & -    & -    & -     & -    & -    & -    & -    & -    & -    & -    \\
31 & 5.97 & 0.15 & 5.06 & 0.14 & 32.02 & 0.90 & 0.29 & 0.01 & 6.15 & 0.36 & 1.89 & 0.11 \\
32 & 6.12 & 0.28 & 4.43 & 0.24 & 28.72 & 1.54 & 0.90 & 0.05 & 6.98 & 0.42 & 6.72 & 0.40 \\
33 & 3.05 & 0.11 & 2.72 & 0.11 & 8.81  & 0.37 & 0.08 & 0.01 & 10.18& 1.62 & 0.84 & 0.13 \\
34 & 0.44 & 0.07 & 2.03 & 0.38 & 0.96  & 0.18 & -    & -    & -    & -    & -    & -    \\
35 & -    & -    & -    & -    & -     & -    & -    & -    & -    & -    & -    & -    \\
36 & 0.55 & 0.02 & 1.94 & 0.09 & 1.13  & 0.05 & -    & -    & -    & -    & -    & -    \\
37 & 8.79 & 0.23 & 4.83 & 0.15 & 44.98 & 1.36 & 0.41 & 0.02 & 5.52 & 0.32 & 2.40 & 0.14 \\
\hline
\multicolumn{12}{l}{A `-' means that no emission above the noise level was detected. } \\
\end{tabular}
\end{table*}

\section{Column densities and abundances}

Tables\,\ref{coldens} and \ref{tab:abund} present
the obtained column densities and abundances for each molecular species at each core.

\begin{table*}[h!]
\centering
\caption{Column densities ($\times 10^{12}$ cm$^{-2}$).}
\label{coldens}
\begin{tabular}{lcccccccccc}
\hline
\hline
Core & N(HC$_3$N) & err. & N(H$^{13}$CN) & err. & N(HN$^{13}$C) & err. & N(H$^{13}$CO$^+$) & err. & N(HC$^{17}$O$^+$) & err. \\
\hline
1  & - & - & 41.20 & 2.07 & 2.43 & 0.10 & 5.86 & 0.11 & 1.33 & 0.14 \\
2  & 494.00 & 80.80 & - & - & 11.90 & 0.95 & 16.20 & 0.76 & 5.62 & 0.34 \\
3  & 96.50 & 18.50 & - & - & 5.27 & 0.24 & 10.30 & 0.35 & 3.07 & 0.26 \\
4  & 212.00 & 9.33 & - & - & 11.20 & 0.27 & 4.68 & 0.17 & - & - \\
5  & - & - & - & - & 0.29 & 0.12 & 3.13 & 0.57 & 0.66 & 0.14 \\
6  & 128.00 & 15.60 & - & - & 11.00 & 0.84 & 2.18 & 0.20 & 6.87 & 0.25 \\
7  & 490.00 & 48.90 & - & - & 18.90 & 0.78 & 10.00 & 0.71 & - & - \\
8  & 46.10 & 9.55 & - & - & 0.97 & 0.11 & 2.99 & 0.19 & - & - \\
9  & 349.00 & 36.80 & - & - & 4.03 & 0.23 & 12.70 & 0.29 & - & - \\
10 & - & - & - & - & 0.47 & 0.39 & 4.94 & 0.25 & 1.08 & 0.16 \\
11 & 360.00 & 25.20 & - & - & 6.81 & 0.21 & 10.70 & 0.31 & - & - \\
12 & - & - & - & - & 0.90 & 0.05 & 4.57 & 0.10 & 0.51 & 0.09 \\
13 & 190.00 & 17.90 & - & - & 1.97 & 0.09 & 10.10 & 0.21 & 2.07 & 0.16 \\
14 & 1010.00 & 135.00 & - & - & 19.50 & 1.28 & 15.90 & 0.68 & 10.40 & 0.55 \\
15 & 40.30 & 2.86 & - & - & 1.53 & 0.09 & 4.81 & 0.09 & 1.19 & 0.13 \\
16 & 53.50 & 6.68 & - & - & 2.93 & 0.13 & 5.08 & 0.13 & 1.20 & 0.19 \\
17 & 82.00 & 5.97 & - & - & 1.47 & 0.24 & 5.99 & 0.11 & 2.10 & 0.13 \\
18 & 98.60 & 7.09 & - & - & 5.23 & 0.17 & 9.82 & 0.27 & 2.02 & 0.12 \\
19 & 315.00 & 31.90 & - & - & 1.21 & 0.09 & 5.63 & 0.10 & 1.17 & 0.09 \\
20 & 834.00 & 44.00 & - & - & 26.00 & 0.38 & 12.50 & 0.81 & 15.90 & 0.99 \\
21 & 158.00 & 17.60 & - & - & 4.94 & 0.24 & 4.96 & 0.15 & 2.06 & 0.20 \\
22 & 591.00 & 82.20 & 178.00 & 4.87 & 25.20 & 0.73 & 23.70 & 0.78 & 5.12 & 0.21 \\
23 & 880.00 & 81.40 & - & - & 6.82 & 0.90 & 9.81 & 0.36 & 5.43 & 0.17 \\
24 & 292.00 & 50.40 & 16.50 & 1.29 & 1.43 & 0.05 & 2.53 & 0.08 & 0.75 & 0.12 \\
25 & 69.50 & 15.70 & - & - & 2.08 & 0.07 & 0.91 & 0.04 & 0.90 & 0.07 \\
26 & 26.10 & 5.17 & - & - & 1.87 & 0.13 & 4.03 & 0.14 & 1.47 & 0.26 \\
27 & - & - & 11.00 & 0.32 & 1.60 & 0.07 & 2.26 & 0.06 & 0.39 & 0.11 \\
28 & 74.50 & 15.70 & - & - & 2.03 & 0.19 & 6.40 & 0.31 & 0.91 & 0.12 \\
29 & 1330.00 & 122.00 & - & - & 44.00 & 1.47 & 9.49 & 0.43 & 11.70 & 0.39 \\
30 & 2100.00 & 416.00 & - & - & 117.00 & 5.86 & - & - & - & - \\
31 & 222.00 & 48.20 & 80.20 & 4.35 & 6.78 & 0.58 & 13.60 & 0.54 & 4.76 & 0.31 \\
32 & 345.00 & 45.70 & - & - & 26.10 & 1.25 & 10.50 & 0.59 & 14.60 & 0.90 \\
33 & 49.80 & 8.21 & - & - & 3.11 & 0.25 & 3.28 & 0.15 & 1.85 & 0.29 \\
34 & - & - & - & - & 0.20 & 0.03 & 0.40 & 0.07 & - & - \\
35 & - & - & - & - & - & - & - & - & - & - \\
36 & - & - & - & - & 0.21 & 0.02 & 0.34 & 0.02 & - & - \\
37 & 1250.00 & 222.00 & 152.00 & 7.07 & 23.70 & 1.08 & 21.00 & 0.80 & 6.66 & 0.41 \\
    \hline
    \end{tabular}
\end{table*}

\begin{table*}[h!]
\centering
\setlength{\tabcolsep}{3.5pt}
\caption{Molecular abundances X  ($\times 10^{-11}$)}
\label{tab:abund}
\begin{tabular}{c cc cc cc cc cc}
\hline
\hline
\multirow{2}{*}{Core} &
\multicolumn{2}{c}{HC$_3$N} &
\multicolumn{2}{c}{H$^{13}$CN} &
\multicolumn{2}{c}{HN$^{13}$C} &
\multicolumn{2}{c}{H$^{13}$CO$^{+}$} &
\multicolumn{2}{c}{HC$^{17}$O$^{+}$} \\
 & X & err. & X & err. & X & err. & X & err. & X & err. \\
\midrule
1  & --     & --     & 35.20 & 4.76  & 2.08  & 0.27  & 5.01  & 0.63  & 1.14  & 0.19  \\
2  & 109.00 & 19.60  & --     & --     & 2.62  & 0.28  & 3.56  & 0.31  & 1.24  & 0.12  \\
3  & 107.00 & 22.60  & --     & --     & 5.82  & 0.59  & 11.40 & 1.10  & 3.39  & 0.42  \\
4  & 439.00 & 38.80  & --     & --     & 23.20 & 1.87  & 9.69  & 0.82  & --     & --     \\
5  & --     & --     & --     & --     & 0.08  & 0.04  & 0.94  & 0.24  & 0.19  & 0.05  \\
6  & 218.00 & 34.60  & --     & --     & 18.70 & 2.38  & 3.71  & 0.50  & 11.70 & 1.26  \\
7  & 189.00 & 22.20  & --     & --     & 7.30  & 0.54  & 3.86  & 0.36  & --     & --     \\
8  & 34.40  & 7.71   & --     & --     & 0.72  & 0.10  & 2.23  & 0.24  & --     & --     \\
9  & 66.70  & 8.06   & --     & --     & 0.77  & 0.06  & 2.43  & 0.15  & --     & --     \\
10 & --     & --     & --     & --     & 0.43  & 0.35  & 4.45  & 0.39  & 0.97  & 0.16  \\
11 & 168.00 & 13.20  & --     & --     & 3.18  & 0.15  & 5.00  & 0.23  & --     & --     \\
12 & --     & --     & --     & --     & 0.42  & 0.04  & 2.15  & 0.15  & 0.24  & 0.04  \\
13 & 65.10  & 6.75   & --     & --     & 0.67  & 0.04  & 3.46  & 0.16  & 0.70  & 0.06  \\
14 & 373.00 & 53.20  & --     & --     & 7.20  & 0.59  & 5.87  & 0.38  & 3.84  & 0.28  \\
15 & 28.60  & 2.65   & --     & --     & 1.09  & 0.09  & 3.41  & 0.21  & 0.84  & 0.11  \\
16 & 55.60  & 7.46   & --     & --     & 3.04  & 0.20  & 5.28  & 0.29  & 1.25  & 0.21  \\
17 & 79.60  & 6.84   & --     & --     & 1.43  & 0.24  & 5.82  & 0.28  & 2.04  & 0.16  \\
18 & 92.10  & 8.54   & --     & --     & 4.89  & 0.32  & 9.18  & 0.59  & 1.89  & 0.16  \\
19 & 139.00 & 17.00  & --     & --     & 0.53  & 0.05  & 2.48  & 0.17  & 0.51  & 0.05  \\
20 & 208.00 & 13.90  & --     & --     & 6.48  & 0.28  & 3.12  & 0.24  & 3.97  & 0.29  \\
21 & 99.40  & 12.70  & --     & --     & 3.11  & 0.24  & 3.12  & 0.21  & 1.30  & 0.15  \\
22 & 367.00 & 57.80  & 111.00 & 8.71  & 15.70 & 1.24  & 14.70 & 1.19  & 3.18  & 0.27  \\
23 & 102.00 & 11.60  & --     & --     & 0.78  & 0.11  & 1.13  & 0.86  & 0.62  & 0.04  \\
24 & 83.00  & 15.70  & 4.69   & 0.52  & 0.40  & 0.03  & 0.71  & 0.06  & 0.21  & 0.03  \\
25 & 65.60  & 15.60  & --     & --     & 1.96  & 0.16  & 0.85  & 0.08  & 0.85  & 0.09  \\
26 & 41.40  & 10.00  & --     & --     & 2.97  & 0.46  & 6.40  & 0.91  & 2.33  & 0.52  \\
27 & --     & --     & 2.72   & 0.23  & 0.39  & 0.03  & 0.55  & 0.04  & 0.09  & 0.03  \\
28 & 665.00 & 155.00 & --     & --     & 1.81  & 0.25  & 5.71  & 0.64  & 0.82  & 0.13  \\
29 & 342.00 & 38.40  & --     & --     & 11.30 & 0.82  & 2.44  & 0.19  & 3.01  & 0.22  \\
30 & 318.00 & 73.90  & --     & --     & 17.70 & 2.33  & --     & --     & --     & --     \\
31 & 87.10  & 22.20  & 31.50  & 4.52  & 2.66  & 0.42  & 5.33  & 0.74  & 1.87  & 0.27  \\
32 & 77.00  & 12.00  & --     & --     & 5.83  & 0.55  & 2.34  & 0.23  & 3.26  & 0.33  \\
33 & 83.70  & 20.80  & --     & --     & 5.23  & 1.06  & 5.51  & 1.06  & 3.11  & 0.75  \\
34 & --     & --     & --     & --     & 0.06  & 0.01  & 0.12  & 0.02  & --     & --     \\
35 & --     & --     & --     & --     & --     & --     & --     & --     & --     & --     \\
36 & --     & --     & --     & --     & 0.09  & 0.01  & 0.15  & 0.01  & --     & --     \\
37 & 356.00 & 74.50  & 43.30  & 5.19  & 6.75  & 0.80  & 5.98  & 0.70  & 1.90  & 0.24  \\ 
\hline
\end{tabular}
\end{table*}

\end{appendix}

\end{document}